\documentclass[multphys,vecphys]{svmult}

\usepackage{makeidx}         
\usepackage{graphicx}        
\usepackage{multicol}        

\makeindex             

\newcommand{\be}{\begin{equation}}
\newcommand{\ee}{\end{equation}}
\newcommand{\ba}{\begin{eqnarray}}
\newcommand{\ea}{\end{eqnarray}}
\newcommand{\mbf}{\mathbf}
\newcommand\tfrac[2]{{\textstyle \frac{#1}{#2}}}
\begin{document}

\title*{Corrections to flat-space particle dynamics arising from space granularity.}
\titlerunning{Dynamical corrections from space granularity}
\author{Luis F. Urrutia}
\institute{Instituto de Ciencias Nucleares, Universidad Nacional
Aut\'onoma de M\'exico, Circuito Exterior, C.U., 04510 M\'exico,
D.F.  \texttt{urrutia@nucleares.unam.mx} }

\maketitle

\begin{abstract}
The construction of effective Hamiltonians describing  corrections
to  flat space particle dynamics arising from the granularity of
space at very short distances is discussed in the framework of an
heuristic approach to the semiclassical limit of loop quantum
gravity. After some general motivation of the subject, a brief
non-specialist introduction to the basic tools employed in the
loop approach is presented. The heuristical semiclassical limit is
subsequently defined  and the application to the case of photons
and spin $1/2$ fermions is described. The resulting modified
Maxwell and Dirac Hamiltonians, leading in particular to Planck
scale corrections in the energy-momentum relations, are presented.
Alternative interpretations of the results and their limitations,
together with other approaches are briefly discussed along the
text. Three topics related to the above methods are reviewed: (1)
The determination of bounds to the Lorentz violating parameters in
the fermionic sector, obtained from clock comparison experiments.
(2) The calculation of radiative corrections in preferred frames
associated to space granularity in the framework of a Yukawa model
for the interactions and (3) The calculation of synchrotron
radiation in the framework of the Myers-Pospelov effective
theories describing Lorentz invariance violations, as well as a
generalized approach to radiation in Planck scale modified
electrodynamics. The above exploratory results show that quantum
gravity phenomenology provides observational guidance in the
construction of quantum gravity theories and opens up the
possibility of probing Planck scale physics.

\end{abstract}

\section{Introduction}
\label{sec:1}

Most theories of gravity suggest that our notion of space-time as
a continuum needs to be revised at  short distances (high
energies) of the order of the Planck length $\ell_P\approx
10^{-33}$ cm (Planck mass, $M_P\approx 10^{19}$ GeV). At these
scales quantum effects should be important and since space-time is
to be considered as a set of dynamical interacting variables,
instead of a mere background where physics occurs,  the quantum
nature of them together with their corresponding fluctuations
could induce modifications to our standard notion of space-time.
The consequences of space-time been considered as a continuum have
been already  successfully probed up to the much lower energies of
$ \approx 10^3$ GeV, corresponding to the standard model of
particle physics.  The sixteen orders of magnitude between our
current experimental arena and the region where such new effects
would become relevant provides in fact plenty of room to look for
modifications of our generally accepted ideas of space-time. We
will refer to this new  possibility as space-time presenting a
granular or foamy structure in the following. Central to this
question is the long time honored problem of finding a consistent
unification of gravity and quantum mechanics, which nowadays is
been actively pursued by loop quantum gravity \cite{REVLQG} and
string theory \cite{REVST}, among others lines of research. At any
rate, a complete formulation of quantum gravity must explain how
the standard notion of space-time at macroscopical length scales
is recovered, thus validating the many successful test of
classical Einstein gravity that have been performed and which have
served to make sure that the correct starting point to construct
the full theory has been taken.

Then a reasonable question to ask is whether or not such short
length (high energy) effects leave any imprint in the dynamics of
particles at standard model energies, which we could be  able to
detect with present day technology and observational
sensibilities.

On one hand, from a purely phenomenological point of view one
would expect some modifications to arise in the same way as
particle propagation properties  change when they move in a
medium, with respect to those in the vacuum. Of course the analogy
of a modified structure of space-time with a propagating medium is
at most very tentative because the idea of a medium presupposes
the existence of something external in which it is embedded. On
the contrary, space-time is the arena where phenomena occur and
there is nothing external to it. Nevertheless it is very plausible
that a drastic change in our description of space-time would
induce modifications in the way we deal with physics at such very
short scales. Here we take the point of view that there is a
remnant of such modifications at standard model energies, which
are described by an effective field theory valid up to scales much
lower than the Planck mass. These corrections manifest themselves
as additional terms contributing to the propagation and
interactions of known particles.

On the other  hand, for a long time it has been taken for granted
that Planck scale phenomena is completely out of reach from
present experiments and/or observations (astrophysical, for
example). Recent investigations show that this is definitely not
the case, thus opening the door to a new area of research called
Quantum Gravity Phenomenology (QGP) that is designed  to use
existing and forthcoming experiments and/or observations to
restrict or constraint competing theories of quantum gravity based
on their predicted imprints at these lower energies \cite{QGPHEN}.
A partial list of references describing such efforts is given in
Ref. \cite{BILLER,analysis}. This is definitely a great advance
over the purely aesthetically criteria that prevailed before. Even
though QGP has made independent advances in restricting  the
different parameters encoding some of the proposed modifications
induced at standard model energies, there still remain  the open
problem of calculating them as a rigorous semiclassical
approximation of a fundamental theory. Only after this gap is
filled one could really be in position to use the observations to
restrict the latter. The work presented here summarizes an
heuristical step towards the estimation of the flat space
semiclassical limit in the realm of LQG, together with further
elaborations upon the results obtained in this way. Also some
closely related topics originating from different approaches are
discussed.

Before going to the details, let us briefly describe some of the
alternative points of view to be found in the literature regarding
the question of whether or not a granular or foamy structure of
space would induce modifications to particle dynamics at standard
model energies. Such modifications have been mainly understood in
terms of the violation of standard Lorentz covariance, either
through the introduction of preferred reference frames or by means
of  the inclusion of an extended or deformed Lorentz relativity.
In the following we will generically  refer to theses possibility
as Lorentz invariance violation (LIV).

A first viewpoint is that modifications that manifest themselves
in the form of LIV, do not arise. In this way covariance under the
standard Lorentz transformations would be perfectly compatible
with a discrete nature of space at Planck scale, which is encoded
in the discrete spectrum of area and volume operators in LGQ, for
example.  This would be analogous to the well known property that
the discrete spectrum of the angular momentum operator does not
preclude the invariance of a system under the continuous rotation
group \cite{ROVSPE,SORKIN}.

A second possibility is that corrections in the form of LIV do in
fact arise. Within this point of view different alternatives have
been also considered, which can be separated in two further
categories:

(i) The first one can be characterized by a phenomenological
parameterization of all possible corrections terms, according to
the dimensionality of the corresponding LIV operators, which are
assumed to arise via a spontaneous Lorentz symmetry breaking of a
more fundamental model like string theory, for example
\cite{STKOSTELECKY}. These vacuum expectation values define a set
of preferred frames, called concordant frames \cite{CONCORDANT},
in which the LIV terms can be  maintained appropriately small when
going from one frame to another via a passive (observer) Lorentz
transformation. This is analogous to the description of atomic
phenomena in the presence of an external magnetic field, where
rotational invariance is broken by active (particle)
transformation. Nevertheless, one can rotate the apparatus and
perform the experiment in the presence of the rotated external
field. The Standard Model Extension (SME)\cite{SME}
 belongs to this class and it has been highly
successful in allowing a unified description of the great amount
of experimental data that has been gathered since 1960
\cite{HUDREV} in relation to the experimental verification of the
isotropy of apace, the transformation properties among inertial
frames and the validity of the discrete transformations C, P and T
as well as its possible combinations \cite{CPTELLIS}. The SME has
been recently generalized to incorporate gravity \cite{SMEGRAV}.
More recently LIV models based upon dimension five operators have
been constructed \cite{MP} and thoroughly analyzed in Refs.
\cite{JLM1,JLM}. Within the  approach (i) we find also direct
extensions of Dirac and Maxwell equations incorporating
modifications which go beyond effective field theories
\cite{CLAUS}. An alternative approach to LIV via spontaneous
symmetry breaking can be found in Ref.\cite{ANDRIANOV}.

(ii) A second broad category of researches which leads to
dynamical corrections  can be distinguished by an effort to obtain
such modifications either from an extension or deformation of the
standard Lorentz relativity principle
\cite{DSRAC,MASMOL,VISSER,GIRLIV,SMOLINR,KOV}, or from a
fundamental description of quantum gravity like, for example: an
effective field theory description of quantum general relativity
\cite{DONAGHUE,DALVIT}, a version of non-critical  string theory
\cite{NCST,NCST1,NCST2} or the approach of loop quantum gravity
\cite{GPED,AMU0,AMU1,AMU2,REVURRU,SEMCLSS}. Alternative ways of
incorporating  such dynamical modifications can be found in
Ref.\cite{PADDY,ALTERN}.

The most immediate way in which the corrections to the dynamics
show up is through modifications to the energy-momentum relations
for the particles \cite{HUET,coleman} \ba\omega_{\pm}
^{2}&=&|{\vec
k}|^{2}\;\pm {\xi \frac{|{\vec k}|^{3}}{E_{QG}}}, \label{phot}\\
E_{R,L}^{2}&=&|{\vec p}|^{2}+m^{2}+{\eta _{R,L}\frac{|{\vec
p}|^{3}}{E_{QG}}},\label{ferm} \ea as shown above for photons and
fermions, respectively.  The connection of such modifications with
gravity, together with the possible astrophysical observation of
the energy-dependent velocity  in high-energy photons arriving
from cosmological sources was suggested in Ref. \cite{ACEMS}. In
fact, the frequency dependent photon velocity \be |{\vec
v}_{\pm}|= \frac{\partial \omega}{\partial |{\vec k}|}= \left(1
\pm {\xi\, \frac{3}{2}\frac{|{\vec k}|}{ E_{QG}}}+ \dots\right)
\ee predicts a time delay $\Delta t $ in the arrival of  two
photons having an energy difference $\Delta E$ and travelling a
distance $L$, given by \be \Delta t\approx \xi\frac{\Delta
E}{E_{QG}}\frac{L}{c}.\ee Here $E_{QG}$ denotes the scale where
quantum gravity effects become relevant, which is usually taken as
the Planck energy.  As emphasized in Ref. \cite{ACEMS}, the tiny
contribution of $\xi/E_{QG}$ can be amplified by selecting large
energy differences or, more effectively, distances $L$ of
cosmological magnitude. It is important to emphasize that this
time delay is a purely kinematical consequence of the modified
dispersion relations.  The orders of magnitude $L\approx10^{10}\,
l.y.,\, \Delta E \approx 20\, MeV,\, E_{QG}=10^{19}\, GeV, \,
\xi\approx 1$, lead to $\Delta t\approx \, 10^{-3}\, s$, which is
within the range of sensitivities $\delta t$ in actual and
forthcoming gamma ray bursts (GRB) observations. In order to
measure such effect it is necessary that $\delta t < \Delta t$.
Present sensitivities allow for estimates of the lower bounds for
$E_{QG}/\xi$ presented in Table \ref{tab:0}.
\begin{table}
\centering \caption{Bounds upon the quantum gravity scale
$E_{QG}/\xi$} \label{tab:0}
\begin{tabular}{lllll}
\hline\noalign{\smallskip} $\qquad  z$ & $ \, \, \delta t (s)$
&$\qquad \quad \quad
\ {\rm Objects}$&$\quad E_{QG}/\xi (GeV) $&$\qquad  \rm{Source}$ \\
\noalign{\smallskip}\hline\noalign{\smallskip} 0.031 $\qquad$ & $
280$ &$\qquad \rm{Markarian 421}$&$\quad  > 4.0\times 10^{16}$&
$\qquad $ Ref.\cite{BILLER} \\
0.0085-3.42 $\qquad$ & $  0.064$ & $\qquad \rm{GRB's \ (BATSE)}
$&$\quad  > 6.9\times 10^{15}$&
$\qquad $ Ref.\cite{ELLISWAVELET}\\
0.3 $\qquad$ & $ 0.031$ & $\qquad \rm{GRB021206 \ (RHESSI)}
$&$\quad  > 1.3\times 10^{17}$&
$\qquad $ Ref.\cite{BOGGS1}\\
\noalign{\smallskip}\hline
\end{tabular}
\end{table}
Future planned observations of GRB at cosmological distances
having well determined red shifts $z$, together with greater
sensitivities ranging form  $10^{-6}\, s$ (RHESSI: Reuben Ramaty
High Energy Solar Spectrometer) to $10^{-7}\, s$ (GLAST: Gamma Ray
Large Area Telescope ) will allow a substantial increase of such
bounds. The fireball model of GRB emission \cite{WAX,VIETRI}
predicts also the generation of $10^{5}-10^{10} \, GeV$ neutrino
bursts which will be detected by observatories like NUBE (Neutrino
Burst Experiment), OWL (Orbiting Wide-angle Light collector
experiment) and EUSO (Extreme Universe Space Observatory), for
example. This will open up the possibility of using such
particles, some times detected in coincidence with the respective
photons, to set further bounds upon the quantum gravity scale
$E_{QG}$.

Very soon after the proposal of Ref.\cite{ACEMS}, the first
derivation of a consistent electrodynamics leading to the
dispersion relations (\ref{phot}) was obtained in the framework of
a loop quantum gravity inspired model \cite{GPED}. The resulting
modified Maxwell equations are \ba \nabla \cdot {\vec E}=0, \quad
\partial_t {\vec E}= - \nabla\times{\vec B} + 2 \,{\xi \,
\ell_P} \, \nabla^2\, {\vec B} ,\\
\nabla \cdot {\vec B}=0, \quad \partial_t {\vec B}= +
\nabla\times{\vec E} - 2 \, {\xi \, \ell_P }\, \nabla^2\, {\vec
E},  \ea leading to the energy-momentum relation \be\\
 \omega_{\pm}=|{\vec k}| \left(1 \mp 2\,
{\xi\, \ell_P}\, |{\vec k}| \right). \ee

Additional bounds upon the  parameters describing the quantum
gravity induced  modifications have been obtained by incorporating
the dynamics through the SME or similar constructions. In
particular, the topics of Lorentz and CPT violations have been
thoroughly studied in low energy physics via theory and
experiments related to:  Penning traps, clock comparison
measurements, hydrogen-antihydrogen studies, spin polarized
dispersion and muon experiments, among others subjects
\cite{BLUHM}.

Finally we mention the use of astrophysical phenomena to discuss
such modified theories. Distinguished examples are the bounds
imposed by polarization measurements from astrophysical sources
\cite{gleiser}, the study of ultra high energy physics processes,
among them cosmic rays, \cite{KIFUNE,ACPIRAN,AMC4,ALFARO,MAJOR}
and the consequences of the detected synchrotron radiation from
the Crab nebulae, as well as that from other objects
\cite{JLM1,JLM,UM,UM1}. For recent reviews about such topics see
Ref.\cite{REVIEWS}.

This contribution summarizes the work carried over in
collaborations with different colleagues and it is organized as
follows. Section \ref{section:2} contains a non-specialist review
of some basic elements of LQG to be subsequently  employed in the
estimation of the induced dynamical modifications. In section
\ref{section:3} we describe the general features involved in the
heuristical calculation  of the effective photon and fermion
Hamiltonians incorporating Planck scale corrections induced by
LQG. These Hamiltonians are presented in subsection 3.3 together
with some related comments. The remaining Section \ref{section:4}
is devoted to selected phenomenological applications of models
presenting Planck scale dynamical corrections. Using existing data
from clock-comparison experiments, in subsection 4.1 we obtain
stringent bounds upon combinations of parameters appearing in the
effective fermionic Hamiltonian previously derived. The
incorporation of radiative corrections to the description of LIV
effects in preferred  frames associated to space granularity is
discussed in subsection 4.2, leading to severe fine-tuning and
naturalness problems. Subsection 4.3 contains the discussion of
synchrotron radiation in LIV electrodynamics, emphasizing the
model of Ref. \cite{MP}, which is phenomenologically formulated as
a theory parameterizing LIV with dimension five operators.
Finally, a unified description of  radiation  in Planck scale
modified electrodynamics including  different models is presented
in subsection \ref{ss102}.

\section{Basic elements from Loop Quantum Gravity (LQG) }

\label{section:2}

Detailed reviews of LQG can be found in Refs.\cite{REVLQG,PEREZ}.
Here we just indicate some of the basic features that will be
relevant in our heuristic calculation and present an intuitive,
non-specialist introduction to this approach. LQG turns out to be
the formulation of Einstein gravity as a Hamiltonian gauge theory
with additional constraints, written in terms of non-local gauge
covariant, diffeomorphisms invariant quantities. It is formulated
in four dimensions and the matter couplings are obtained by
rewriting the standard ones in terms of the new variables. This
background independent, non-perturbative theory has allowed
substantial progress in the old problem of producing a consistent
quantum description of Einstein gravity. It has successfully dealt
with traditional problems like a macroscopic account of the black
hole entropy \cite{CANBH} and the construction of non-singular
cosmological models \cite{NONSCOSM}, among other topics. One of
its most notable predictions is the property that area and volume
operators are quantized in terms of the corresponding powers of
$\ell_P$, thus signaling a granular structure of space at short
distances \cite{volumeop}.  It is precisely the possible
consequences that such granularity  may induce  at energy scales
of the standard model what is to be explored in this work. This
question is intimately related to the still open problem of the
semiclassical limit in LQG:  how does one recover the continuous
metric description of space-time starting from the quantum version
of it?.

\subsection{The passage to the new variables}

LQG is formulated in terms of the Ashtekar-Barbero variables that
arise as a canonical transformation from the usual ADM variables.
Here we summarize the main steps leading to this choice and
closely follow Ref.\cite{PEREZ}.

Let us start from the standard  Einstein-Hilbert action for
gravity in the signature $(-,+,+,+)$ \be S[g_{\mu \nu
}]=\frac{1}{2\kappa }\int \left( d^{4}x\right) \;\sqrt{-\det
(g_{\mu \nu })}\; {}^{(4)}R, \quad \kappa =\frac{8\pi G}{c^{3}},
\label{EIACTION} \ee where ${}^{(4)}{R}$ is the four dimensional
Ricci scalar. The ADM variables result from a foliation of
spacetime  in {3D} surfaces ${\Sigma}$ of constant parameter
{$t$}, described by coordinates {${x^a, a=1,2,3}$}, together with
the parameterization of the ten variables {${g_{\mu\nu}}$} in
terms of the six components on the induced three-metric
{${q_{ab}}$} on {$\Sigma$}, plus the three components of the shift
vector {${N^a}$} and the lapse function {${N}$}. In terms of these
variables the four dimensional metric is \be{
g_{tt}=q_{cd}N^{c}N^{d}-N^{2}, \quad
g_{ta}=g_{at}=q_{ac}N^{c},\quad  g_{ab}=q_{ab} }\ee and  the
action (\ref{EIACTION}) reads \be
S[q_{ab},\;N^{a},\;N]=\frac{1}{2\kappa }\int dt\int \left( d^{3}x\right) \;%
\sqrt{\det (q_{ab})}\;N\;\left( ^{(3)}R+K_{ab}K^{ab}-\left(
K_{a}^{a}\right) ^{2}\right), \ee where all the indices are now
lowered or raised by $q_{ab}$ and its inverse $q^{cd}$. Here
$^{(3)}R$ is the Ricci  scalar of the manifold ${\Sigma}$. The
information about the velocities $\dot{q}_{ab}$ is contained in
the extrinsic curvature \be K_{ab}=\frac{1}{2N}\left(
\dot{q}_{ab}-\mathcal{L}_{\vec{N}}\;q_{ab}\right) =K_{ba}, \ee
where $\mathcal{L}_{\vec N}$ denotes de Lie derivative along the
vector ${N^a}$. Introducing the canonical momenta  \ba
&&\frac{1}{\kappa }\Pi ^{ab}=\frac{\delta S}{\delta \dot{q}_{ab}}=\frac{1}{%
\kappa }\left( K^{ab}-q^{ab}K_{c}^{c}\right) \sqrt{\det
(q_{ab})},\\
&&\Pi= \frac{\delta S}{\delta \dot{N}}=0, \qquad  \Pi_a=
\frac{\delta S}{\delta \dot{N}^a}=0 \ea and making a Legendre
transformation one finds {\begin{eqnarray} &&{S[q_{ab},\;\Pi
^{ab},\;\;N^{a},\;N] =\frac{1}{2\kappa }\int dt\int \left(
d^{3}x\right) \left[ \Pi
^{ab}\,\dot{q}_{ab}-N^{a}H_{a}-N\;H\right]}.\nonumber \\
\label{ACTION1}
\end{eqnarray}}
The above action shows that gravity is a fully constrained theory
with zero canonical Hamiltonian. The invariance under local
diffeomorphisms in $\Sigma$ is generated by the constraints $H_a$,
while the dynamics is generated  by the Hamiltonian constraint
$H$. These are first class constraints which can be written as
explicit functions of the canonical variables. The shift and the
lapse functions turn out to be Lagrange multipliers.

The symplectic structure corresponding to (\ref{ACTION1}) is given
by the Poisson brackets \be \left\{ \Pi ^{ab}({\vec
x}),\;q_{cd}({\vec y})\right\} =2\kappa \delta _{(c}^{a}\delta
_{d)}^{b}\delta ^{(3)}({\vec x},{\vec y}), \ee with the remaining
ones being zero.

The second step towards the construction of the  new variables is
to introduce a non-abelian $SO(3)\,\approx\, SU(2)$  formulation
of the action (\ref{ACTION1}). To this end one rewrites the
three-metric in terms of a triad $e_a^i$ such that \be
q_{ab}=e_{a}^{i}e_{b}^{j}\delta _{ij}, \label{TRIAD}\ee where the
indices $i, j=1,2,3,$ transform under local rotations.

The canonical variables arising in this step are the densitisized
(weight $+1$) triad \be E_{i}^{a}=\frac{1}{2}\epsilon
^{abc}\epsilon _{ijk}e_{b}^{j}e_{c}^{k}, \label{DT} \ee together
with the projected extrinsic curvature \be
K_{a}^{i}=K_{ab}E_{j}^{b}\delta ^{ij}. \ee In terms of them we can
rewrite \be \det(q_{ab})=\det(E^a_i), \qquad \dot{q}_{ab}\Pi
^{ab}=2\dot{K}_{a}^{i}\,E_{i}^{a},\ee and the action is
\begin{eqnarray} S\left[ E_{i}^{a},\;K_{b}^{j},N^{a},N, N^i
\right]}={\frac{1}{\kappa }\int dt\int \left( d^{3}x\right) \left[
E_{i}^{a}\,\dot{K}_{a}^{i}-N^{a}H_{a}(E,K)\right.&& \nonumber\\
\left. -N\,H(E,K)-N^{i}G_{i}(E,K)\right], &&
\end{eqnarray}
The introduction of an extra  $SO(3)$ gauge freedom in
(\ref{TRIAD}) requires the additional constraints \be
G_{i}(E,K)=\epsilon _{ijk}E^{aj}K_{a}^{k}, \ee which turn out to
be just the non-abelian Gauss law.

The final step to the Ashtekar-Barbero variables \cite{ASHTBEARB}
starts from the recognition that there is a natural
${{SO(3)\approx SU(2)}}$ connection ${\Gamma _{a}^{i}}$ compatible
with
the triad, such that%
\be
\partial _{\lbrack a}e_{b]}^{i}+\epsilon _{\;jk}^{i}\Gamma _{\lbrack
a}^{j}e_{b]}^{k}=0.\ee Moreover, the triad introduced in Eq.(\ref
{TRIAD}) can be expressed in terms of the densitized triad as \be
e_{a}^{i}=\frac{1}{2\sqrt{|\det (E)|}}\,\, \epsilon _{abc}\epsilon
^{ijk}E_{j}^{b}E_{k}^{c},\qquad e_{i}^{a}=\frac{sgn(\det
(E))}{\sqrt{|\det (E)|}}\,\, E_{i}^{a}. \ee The final
Ashtekar-Barbero connection, which allows to write the constraints
in a simpler form,  is \be A_{a}^{i}=\Gamma _{a}^{i}+\gamma
K_{a}^{i}. \label{ABCONN}\ee Here $\gamma$ is the Immirzi
parameter \cite{INMIRZI} and  the corresponding field strength
two-form is \be {{F^i=d A^i+ \epsilon^i{}_{jk} A^j\wedge A^k}},
\qquad F^{ij}_{ab}=\epsilon^{ij}{}_k\, F^k_{ab}.\ee The final
action results
\begin{eqnarray} S\left[ E_{i}^{a},\;A_{b}^{j},N^{a},N, N^i
\right]}={\frac{1}{\kappa }\int dt\int \left( d^{3}x\right) \left[
E_{i}^{a}\,\dot{A}_{a}^{i}-N^{a}H_{a}(E,A)\right.&& \nonumber\\
\left. -N\,H(E,A)-N^{i}G_{i}(E,A)\right], && \label{ABACTION}
\end{eqnarray}
with the canonical variables satisfying the non-zero Poisson
brackets \be \{  E_j^b ({\vec y}), A_a^i({\vec x}) \}= \kappa\,
\gamma \, \delta^b_a \, \delta^i_j \, \delta^{(3)}({\vec x}, {\vec
y}).\label{FINPB}\ee The explicit form of the constraints are \ba
&&G_i=D_a\ E^a_i, \qquad
H_b=E^a_i\, F^i_{ab}-(1+\gamma^2) K_b^i\,G_i, \nonumber \\
&&H=\frac{1}{\sqrt{|\det(E)|}}\left(F_{ab}^{ij}-2 (1+\gamma^2)
K^i_{[a}\, K^j_{b]} \right) \, E^a_i \, E^b_j. \ea The choice
$\gamma=i$ corresponds to the  original Ashtekar variables, which
define a complexified version of Einstein gravity together with a
very simple structure for the Hamiltonian constraint. The price to
be paid is the need  to incorporate  some reality conditions in
order to recover the corresponding real formulation.

Following  the standard Dirac quantization procedure  arising from
the action (\ref{ABACTION}), together with  the Poisson brackets
(\ref{FINPB}) in the coordinate space defined by the connection
$A_a^i$, we promote the canonical variables to operators in such a
way  that \be {\hat A}_a^i \Psi[A]={\ A}_a^i \Psi[A], \qquad {\hat
E}^a_i \Psi[A]=-i\hbar\, \kappa\,\gamma
\frac{\delta}{\delta{A}_a^i} \Psi[A] \ee and impose the operator
version of the constraints as null conditions upon the wave
function.
\subsection{Holonomies and fluxes}
\begin{figure}
\centering {\includegraphics[width=5.5cm]{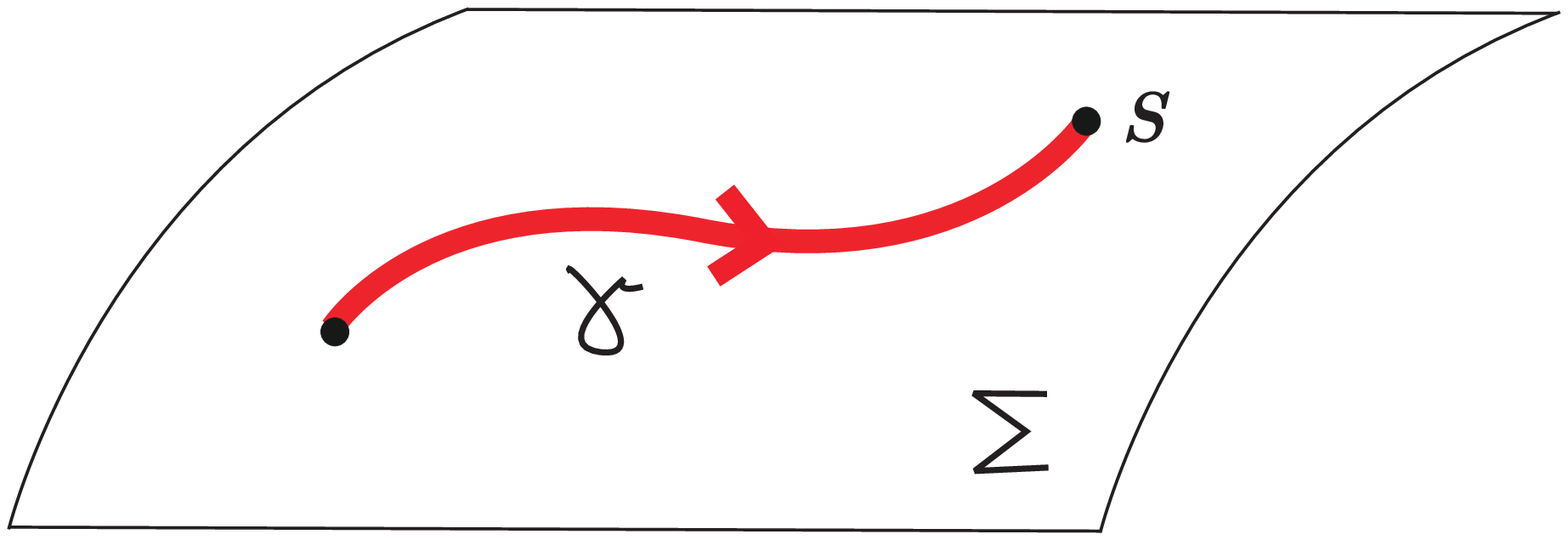}}
{\includegraphics[width=5.5cm]{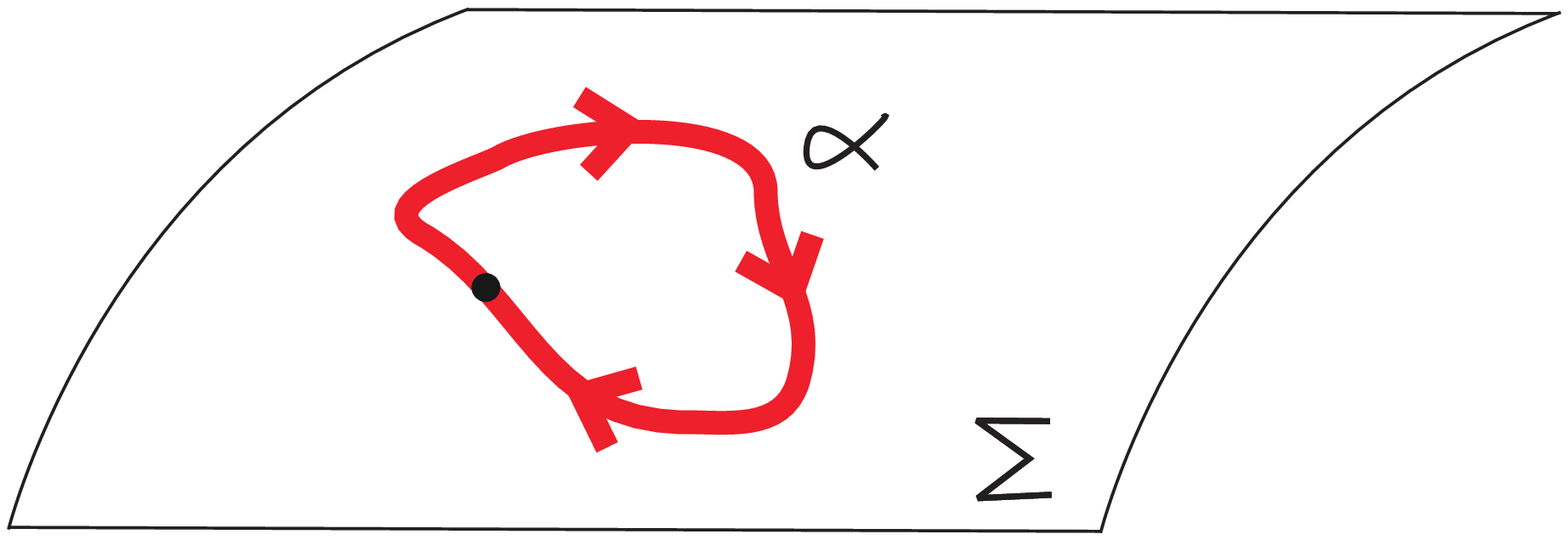}} \caption{Open and
closed curves in $\Sigma$ } \label{fig:0}
\end{figure}

Next we describe the fundamental operator variables employed to
formulate  LQG  and which supersede the quantum version of the
previous canonical variables (\ref{DT}) and (\ref{ABCONN}). The
basic support for the operators in LQG are open or closed curves
in $\Sigma$, as depicted in Fig. \ref{fig:0}, respectively,
together with two-dimensional surfaces. These allow to define
appropriately smeared versions of the operators (\ref{DT}) and
(\ref{ABCONN}).
The $SU(2)$ connection operators $A^i_a({\vec x})$ are replaced by
holonomies (or parallel transport matrices) $U$ along  open or
closed curves in the thee-manifold $\Sigma$, defined as \be
U(\gamma,A)={\cal P}\, {\rm exp} \,i \int_{0,\,\gamma}^s
A^i{}_a(\gamma(s'))\, \sigma_i\, \frac{dx^a}{ds'}\,ds'\,\,\, \in
SU(2). \label{HOLO}\ee The notation is the following:  the Pauli
matrices $\sigma^i$ are the generators of the  $SU(2)$ Lie
algebra. The open curve $\gamma$ is parameterized by $s'$ with $0<
s' < s$ and $\cal P $ denotes the standard path-ordered product.
The holonomy, which is a multiplicative operator, transforms
covariantly under $SU(2)$ gauge transformations at the end points,
but still it is not invariant under diffeomorphisms.

The canonically conjugated momenta operators ${\hat E}^a_i$ are
replaced by their corresponding fluxes over  surfaces $S$ in
$\Sigma$ \be {\hat E}_i(S)=-i\hbar \int_S \,d\sigma^1\, d\sigma^2
\,n_a({\vec \sigma})\, \frac{\delta}{%
\delta {\ A}_a^i(x({\vec \sigma}))}, \quad  n_a=\epsilon_{abc}
\frac{\partial x^b}{\partial\sigma^1}\frac{\partial x^c}{%
\partial\sigma^2}.
\label{FLUX} \ee
 Here ${\vec \sigma}=(\sigma^1, \sigma^2)$ are the intrinsic coordinates of the
surface and $n_a$  is its normal vector. The above definitions
$(\ref{HOLO}) $ and $(\ref{FLUX})$ allow to calculate the
corresponding commutator, which turns out to be a non-canonical
one.

\begin{figure}
\centering
\includegraphics[height=7cm]{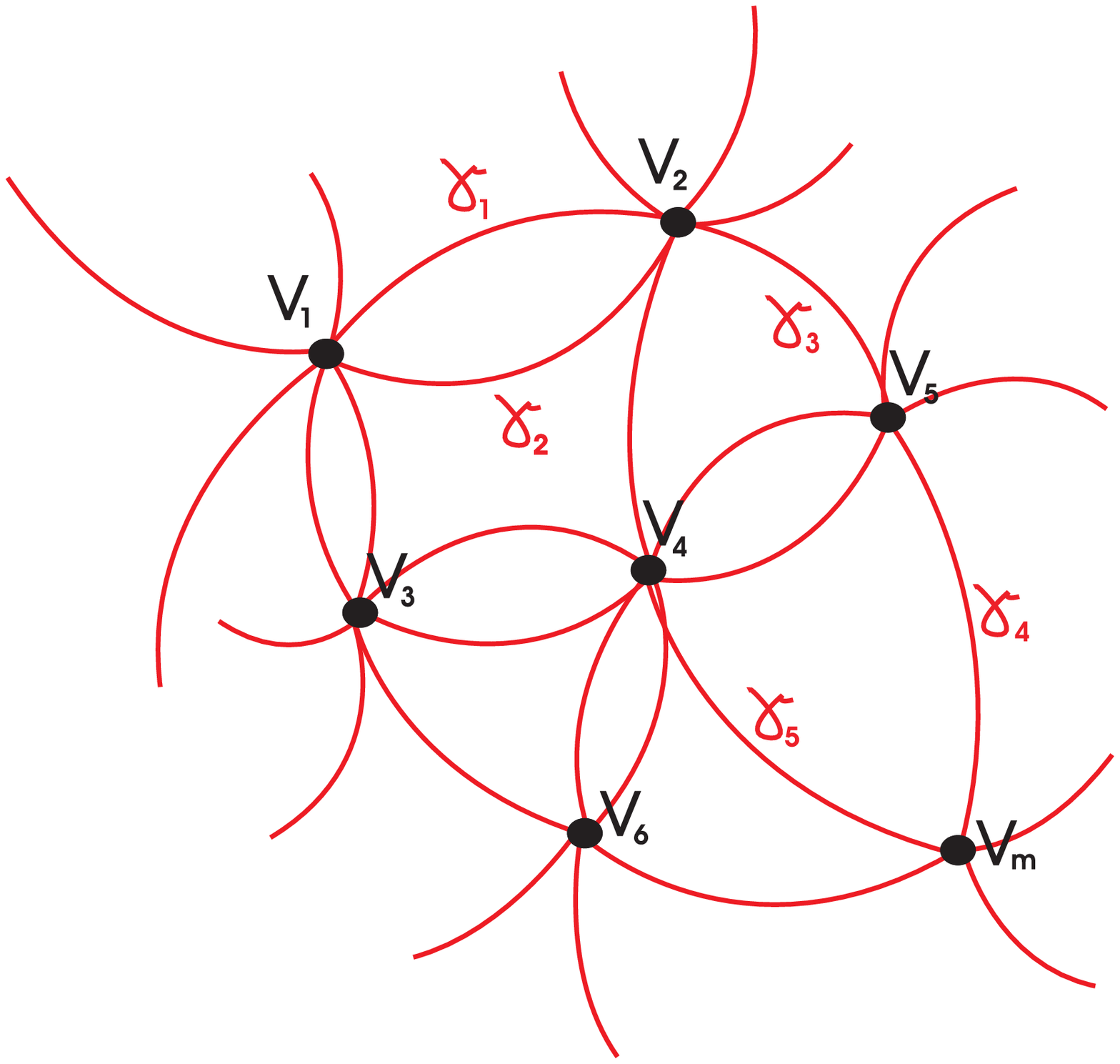}
\caption{Piece of a  graph ${\Gamma=\{\vec{V}_1,\dots,\vec{V}_n;
\gamma_{1}, \dots,\gamma_m\}} $ } \label{fig:1}
\end{figure}

Quantum states $\Psi_{\Gamma,f} (A)$ are  represented by
functionals of generalized connections ($SU(2)$ group elements)
defined over graphs in $\Sigma$, which  are called cylindrical
functions and define the kinematical space of the problem. As
represented in Fig. \ref{fig:1}, a graph
${\Gamma=\{\vec{V}_1,\dots,\vec{V}_n; \gamma_{1},
\dots,\gamma_m\}} $ is a set of points
$\{\vec{V}_1,\dots,\vec{V}_n\} \in \Sigma $, called the vertices,
joined by curves $\gamma_{1}, \dots,\gamma_m$, called the edges of
the graph. The number of edges attached to a vertex is called the
valence of the vertex.
To each edge of the graph we associate a group element $U(\gamma,
A)$  labelled by  an irreducible representation of $SU(2)$ and
consider a function $f$ which is a map from the direct product of
group representations to the complex numbers, so that \be
{\Psi_{\Gamma,f}(A)=f(U(\gamma_1, A), \dots, U(\gamma_m, A) )}\
\in \ \rm{Kinematical \ Space}. \ee For a given graph $\Gamma$ we
define the scalar product of two associated functions $f$ and $g$
as \be<\Psi_{\Gamma, f}|\Psi_{\Gamma, g}>=\int (\prod_{\gamma_i}
dU_i)f^{\ast}(U_1\dots U_m)g(U_1\dots U_m), \label{QUANTST}\ee
where we have denoted $U_i:=U(\gamma_i, A)$. Here $dU_i$ is the
corresponding Haar measure for the group $SU(2)$.

There is still  a long way to construct the physical space. To
this end one has to subsequently impose $SU(2)$ gauge invariance,
three dimensional diffeomorphism invariance and finally  the
Hamiltonian constraint. The first two conditions can be explicitly
accomplished. The first step is implemented by an adequate choice
of the functions $f$ in terms of the so-called intertwiners at
each vertex. These are invariant tensors in $SU(2)$ which map the
product of representations at each vertex into gauge invariant
expressions. In these way the so called  spin network states are
defined and it is possible to prove that  they form an orthonormal
basis for the $SU(2)$ gauge invariant kinematical space. The
second step is performed by going from loops to knots, thus
defining equivalent classes of loops under diffeomorphisms.

\begin{figure}
\centering
\includegraphics[height=4cm]{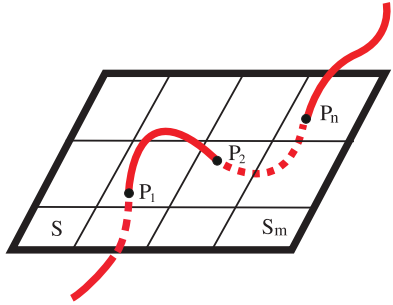}
\caption{Conditions for the action of the smeared momentum
operator} \label{fig:4}
\end{figure}

The action of the operators (\ref{HOLO}) and (\ref{FLUX}) upon
quantum states (\ref{QUANTST}) is defined by \ba
(U^A_B(A,\gamma)\Psi)[A]&=& U^A_B(A,\gamma)\times\Psi[A],\nonumber
\\
{\hat E}_i(S) U(A, \gamma)&=&\sum_k \pm i\hbar\,
U(A,\gamma_1^{P_k})\, \sigma_i \, U(A,\gamma_2^{P_k}). \ea Thus,
holonomies act as multiplicative operators, while the action of
the smeared conjugated momentum depends upon a given edge $\gamma$
of the graph crossing or not the associated surface $S$ at the
points $P_k$. For each intersection $P_k$ the path $\gamma$ is
separated in two pieces $\gamma_1^{P_k}$ and $\gamma_2^{P_k}$ such
that $\gamma=\gamma_1^{P_k}\,\circ \, \gamma_2^{P_k} $. The action
at that intersection  insert the generator $\sigma_i$ between the
holonomies corresponding to each of these paths. This is shown
schematically in Fig. (\ref{fig:4}).
To close this summary we mention the area operator ${\hat A}$
which is defined in the following way \be {\hat A}= \lim_{n
\rightarrow \infty} \sum_{n}\sqrt{{\hat{E}_i(S_n)\hat{E}_i(S_n)}},
\ee
 by partitioning the associated  surface $S$. A similar
 construction is made for the volume operator ${\hat V}$. These
 are well defined partial observables
(hermitian operators in the $SU(2)$ gauge invariant  kinematical
space) with discrete eigenvalues and eigenfunctions given by the
spin network states. An important property to be used in the
following is that only the nodes contribute to the action of the
volume operator upon states defined in a graph.
\subsection{Elements of Thiemann regularization}
\begin{figure}
\centering
\includegraphics[height=6cm]{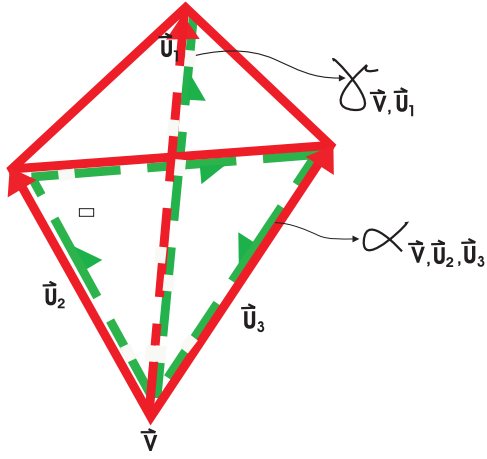}
\caption{Basic tetrahedron and paths in Thiemann regularization}
\label{fig:3}
\end{figure}

Thiemann has proposed a general regularization scheme that
produces a consistent mathematical definition for the operators
entering in the  description of loop quantum gravity
\cite{THIEMREG,THIEMREG1}. Such regularized operators act upon
states which are functions of generalized connections defined over
graphs. The regularization procedure is based upon a triangulation
of space which is adapted to each graph. This means that the space
surrounding any vertex of $\Gamma$ is filled with tetrahedra
$\Delta $ having only one vertex in common with the graph (called
the basepoint ${\vec V}(\Delta)$) plus segments ${\vec
U}_I(\Delta), I=1,2,3$, starting at ${\vec V} $  and directed
along the edges of the graph. For a vertex of valence greater than
three one must consider all possible combination of three  edges
to build the tetrahedra.  In the regions not including the
vertices of $\Gamma$ the choice of tetrahedra is arbitrary and the
results are independent of it.  The open path along the segment
${\vec U}_I$ is denoted by $\gamma_{{\vec V}, {\vec U}_I}$. The
arc connecting the end points of ${\vec U}_I(\Delta)$ and ${\vec
U}_J(\Delta)$ is denoted by $a_{IJ}(\Delta)$ and the loop
$\alpha_{{\vec V}, {\vec U}_I, {\vec U}_J}:= \gamma_{{\vec V},
{\vec U}_I} \, \circ \,\, a_{IJ}\, \circ \,\, \gamma^{-1}_{{\vec
V}, {\vec U}_J}:=\alpha_{IJ}(\Delta)$ can be formed. This
construction is illustrated in the basic tetrahedron depicted in
Fig. (\ref{fig:3}).

A first step in the implementation of the  regularization method
is to express  each connection and field strength in terms of an
adequate holonomy. For example, using the path ${\gamma_{{\mathbf
{\vec V}}, {\mathbf{\vec U_1}}}}$, of coordinate length
${\epsilon}$, we can incorporate a connection $A_a({\vec x})$ into
the corresponding holonomy  via the expansion \be
U(\gamma_{{\mathbf {\vec V}}, {\mathbf {\vec U_1}}}, A )=
{1+\epsilon U_1^a A_a(x)+O(\epsilon^2)}. \label{PATH}\ee An
analogous calculation for  the closed triangular path ${{\alpha_{
{\mathbf {\vec V}}, {\mathbf {\vec U}_2}, {\mathbf {\vec U}_3}
}}}$ leads to the following expression incorporating the curvature
\be U( \alpha_{ {\mathbf{\vec V}}, {\mathbf {\vec U}_2}, {\mathbf
{\vec U}_3} }, A )={1+ \frac{1}{2}\epsilon^2 U_2^a \, U_3^b\
F_{ab} + O(\epsilon^3)}.\label{LOOP} \ee

A second ingredient of the method is the basic identity
\cite{THIEMREG} \be \frac{E^a_i({\vec x}) E^b_j({\vec x})
\epsilon^{ijk}}{\sqrt{det(E({\vec x}))}}=\frac{2}{\kappa}
\epsilon^{abc} \{ A_c^k , V(R) \}_{PB}, \quad V=\int_R\,
(d^3x)\sqrt{det(E^{ai})}, \label{TID}\ee
 where $R$ is a region such ${\vec x}\in R$. In this way
the volume $V$, which will be promoted to the operator level once
the quantization is performed, is introduced in the
regularization.

To fix ideas let us look at the  first piece of the smeared
Hamiltonian constraint
 \be{{{\cal H}_1=\int (d^3x)\, N(x) \, \frac{F^{k}_{ab}\epsilon^{ij}{}_k \,E^a_i E^b_j}{\sqrt{det(E)}}}
{=\int(d^3x) N(x) \, \epsilon^{abc}F^i_{ab}\, \{ A_{ci}, V \})
}},\ee where we have used Eq.(\ref{TID}) in the second step. Now,
substituting  the expressions  (\ref{PATH}) and (\ref{LOOP}) for
${{A_a,\, F_{ab}} \in SU(2)}$, respectively,  and replacing the
integral with a sum, we can write \ba && {{{\cal
H}_1=\lim_{\epsilon \rightarrow 0} \sum_m {\epsilon^3 }\, N_m \,
\epsilon^{IJK}\frac{1}{\epsilon^3}Tr\left(\frac{}{}\left(
U(\alpha_{{\vec V}_m, {\vec U}_I, {\vec U}_J}, A)-
U^{-1}(\alpha_{{{\vec V}_m}, {{\vec U}_I}, {\mathbf{\vec U}_J}},
A)\right)\,\right.}\nonumber} \\
&&\qquad\qquad \qquad {{\left.\times \, {U^{-1}(\gamma_{{{\vec
V}_m}, {{\vec U}}_K}, A )}\, \{ U(\gamma_{ {\vec V}_m, {\vec U}_K
}, A ), V(R_m) \} \frac{}{}\right)}}. \label{REGHAMC} \ea  The
crucial fact is that the divergent contribution arising from
$1/\sqrt{det(E)}$, which goes like $1/\epsilon^3$, is cancelled by
the correspondent factors contained in expansion of the
holonomies. In this way Eq. (\ref{REGHAMC}) leads to a well
defined regulated expression. Furthermore, when the correspondent
quantities are quantized, the volume operator ${\hat V}$ is
naturally introduced. This has the important consequence that the
action of such operators upon  wave functions defined on a graph
get contributions only from the nodes of the graph.

Another illustrative example of this regularization is the
magnetic sector of the electromagnetic Hamiltonian
\be
 H^B_{{\rm Maxwell}} = \int_{\Sigma} d^3x \,
\frac{q_{ab}}{2Q^2\sqrt{\det q}} \underline B ^a \underline B^b,
\ee which is translated into the operator expression
\begin{eqnarray}
{ {\hat H}^{B}} &=&{\frac{1}{2\,\ell_P^4\,Q^{2}}\sum_{{\vec V}\in
V(\Gamma
)}\,\, \sum_{{\vec V}(\Delta )={\vec V}({\Delta'})={\vec V}}\epsilon ^{JKL}\, \epsilon ^{MNP}\times} \nonumber \\
&&{\times}  {\hat w}_{i L \Delta}\,\, {\hat w}_{i P \Delta'}
 \, \,
 \left({\underline{h}}_{\alpha _{JK}(\Delta )}-1 \right)
\,\left({ \underline{h}}_{\alpha _{MN}(\Delta')}-1\right),
\label{REGHAMB}
\end{eqnarray}
where \be{\hat w}_{k I \Delta}= Tr\left( \tau_k h_{{\vec
U}_I(\Delta)} \left[ h^{-1}_{{\vec U}_I(\Delta)},\sqrt{{\hat
V}_v}\right] \right).\label{GRAVOP}
 \ee Here we  have simplified the notation by introducing
\be h_{{\vec U}_I(\Delta)}=U(\gamma_{{{\mathbf {\vec V}}_\Delta},
{\mathbf {\vec U}}_I}, A ), \qquad  {\underline{h}} _{\alpha
_{JK}(\Delta )}= U(\alpha_{{\mathbf {\vec V}}_\Delta, {\mathbf
{\vec U}_J}, {\mathbf {\vec U}_K}}, \underline{A}). \ee The
underlined quantities (holonomy and connection) refer to the
electromagnetic sector, and are to be distinguished from the
gravitational ones.

In the following we will separately consider each matter
contribution to the Hamiltonian constraint as the corresponding
Hamiltonian for the respective gravitationally coupled sector.
\section{A kinematical estimation of the semiclassical limit}
\label{section:3}
The construction of states that live in the
physical space of LQG and  which approximate a given geometry at
large distances, while retaining their granular structure at
Planck length scales is still an open problem. This is usually
referred to as the semiclassical limit of LQG and it is actively
under investigation \cite{COHSTATES}.
\subsection{Heuristic characterization of the states}
In order to make some preliminary steps towards the study of the
consequences of space granularity at large scales in the flat
space limit, we take here an heuristical point of view, starting
from the exact and well understood operator version of LQG. The
approximation we introduce consists in defining their action upon
the semiclassical states through some plausible requirements,
without having an explicit construction for such states.  We think
of the semiclassical configuration corresponding to a particular
matter or gauge field, described by operators collectively denoted
by ${\hat F}({\vec x})$, plus gravity  as given by an ensemble of
graphs $\Gamma$, each occurring with probability $P(\Gamma)$. To
each of such graphs we associate a wave function $|\Gamma, {\cal
L}, F \rangle := |\Gamma, S \rangle$ which we assume to be peaked
with respect to the classical matter field configuration $F({\vec
x})$, together with a flat gravitational metric and a zero value
for the gravitational connection at large distances. Not
surprisingly, the semiclassical approximation gives us
simultaneous average information regarding a field together with
its canonically conjugated momentum.  In other words, the
contribution of the gravitational and matter operators inside the
expectation value is estimated as \cite{AMU0,AMU1,AMU2}
\begin{eqnarray}
&&{\langle \Gamma, S|\dots {\hat q}_{ab} \dots|\Gamma,
S\rangle\approx
\delta_{ab}+O\left(\ell_P/\cal L\right)}, \nonumber\\
&&{\langle \Gamma, S|\dots {\hat A}_{ia} \dots|\Gamma,
S\rangle\approx \left( 0+(1/{\cal L})\left(\ell_P/{\cal
L}\right)^{\Upsilon}\right)_{ia}}\, ,\nonumber\\
&&{\langle \Gamma, S|\dots {\hat {\underline{F}}}{\,}_{ab}
\dots|\Gamma, S\rangle  \approx \mu  {\underline{F}}{\,}_{ab}}
\label{EXPV}
\end{eqnarray}
and similarly for any product of operators inside the expectation
value. As an illustration, the gauge field  is taken to be the
electromagnetic field in the following  but an analogous
estimation will hold for any other gauge or matter fields. The
parameter $\Upsilon \geq 0$ is a real number and
${\underline{F}}{\,}_{ab}$ denotes the classical electromagnetic
field strength. We further associate the effective hamiltonian \be
H_{\Gamma}=\langle \Gamma,S\,|{\hat H}_{\Gamma}
 |\, \Gamma,S \rangle
 \label{GEFFHAM}
 \ee
to each graph.

The coarse graining scale ${\cal L}>> \ell_P \, $ of the wave
function is such that a continuous
 flat metric approximation is appropriate for distances $D$ much larger that ${\cal
L}$,
 while the granular structure of space becomes relevant when probing
 distances $D$ smaller that ${\cal
L}$. That is to say, ${\cal L}$ is not a scale of quantum gravity
but rather it is a scale that separates the continuum description
of space from  its discrete characterization that is fully
manifest at the Planck length $\ell_P$, which signals the quantum
behavior. Summarizing, we expect  the following behavior
\begin{eqnarray}
 D \,>>\, {\cal L} \,>> \, \ell_P & : & {\rm \ continuous \ flat \
classical \ geometry, }
\nonumber\\
\ell_P \sim  D \,<<\,  {\cal L} & : & {\rm  manifest \  discrete \
structure \ of \ space, } \nonumber
\end{eqnarray}
for the probe scale $D$. The coarse graining scale ${\cal L}$ is
not provided by our approximation  and has to be  estimated in
each particular case. We will explore some alternatives in
subsection \ref{subsec:param}.

\begin{figure}
\centering
\includegraphics[width=11.5cm]{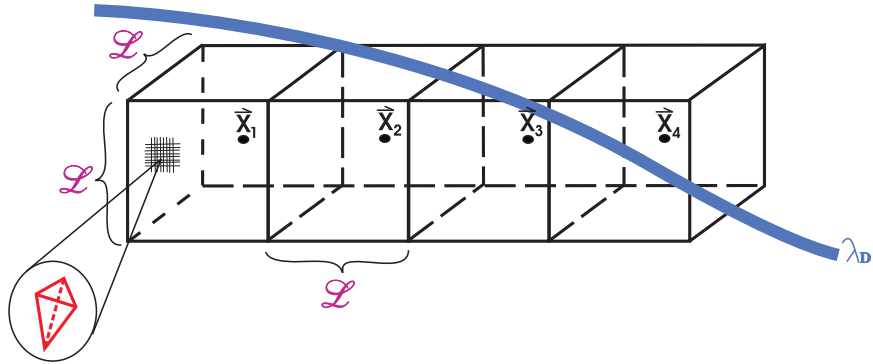}
\caption{Different scales for the model of space considered in the
text} \label{fig:2}
\end{figure}

 As indicated  in Fig. \ref{fig:2}, we think of space as constructed by
 adding boxes of size ${\cal L}^3$, which center represents a given point
$\bf x$ in the
 continuum limit and which contain a large number of vertices of the adapted graphs.
 The matter field $F({\vec x})$, characterized by a de Broglie wave length
 $\lambda$,
 is considered as a slowly varying function within each box ($\lambda >
 {\cal L}$) and  contributes with  its classical value at the center
 of the box, when taking expectation values. The requirement $\lambda >
 {\cal L}$ guarantees that we can describe the flat space dynamics
 in terms of an effective field theory, using the standard
 differential calculus of the continuum. On the other hand,
gravitational variables are rapidly varying inside each box.

The total effective Hamiltonian is defined as an  average over the
graphs $\Gamma$ \be{\rm H}=\sum_{\Gamma} P(\Gamma)\, {\rm
H}_\Gamma. \label{TEFFHAM}\ee
\subsection{The calculation}
We summarize now the method of calculation \cite{AMU0,AMU1,AMU2}.
For each graph $\Gamma$  the
 effective Hamiltonian is defined by (\ref{GEFFHAM}).
For a given vertex $\vec V $ inside the expectation value, one
expands each holonomy or flux operator in powers of the segments
$U^a_I(\Delta))$ of the attached tetrahedra $\Delta$, plus
derivatives of the gauge or matter fields operators.
Schematically, in the case of (\ref{REGHAMB}) this produces
\begin{eqnarray}
{ { H}^B_{\Gamma} }&=&{ \sum_{{{\vec V}\in \, {\vec V}(\Gamma
)}}\,\sum_{{ {\vec V}(\Delta )}={ {\vec V}}}\, \langle { \Gamma ,S
|\left( \frac{}{}
\partial^{a_1}\partial^{a_2}\dots\partial^{a_k}\dots{\hat {\underline
F}}_{p_1q_1}({\vec V})\right)\dots \times
 }}\nonumber\\
&& \times\, \left(\dots \frac{}{} \partial^{a_{m-1}}
\partial^{a_m}{{\hat {\underline F}}_{pq}({\vec V})}\right)
{{{\hat T}}_{{ a_1 }\dots { a_m}}^{{ p_1}\, { q_1}\,...{ p}\,{
q}}({ {\vec V}, {{\vec U}_I}({\Delta), {{\hat A}_{ia}}}},{\hat
V})} \,\, | \Gamma,  S \rangle. \nonumber \\ \label{REGHAM}
\end{eqnarray}
Here ${\hat T}$ contains  gravitational operators (connection and
volume operators, for example) together with contributions
depending on the segments of the adapted triangulation in the
particular graph. Next, according to Fig. \ref{fig:2},  space is
considered to be divided into boxes, each centered at a fixed
point ${\vec{x}}$ and with volume ${\cal L}^{3}\approx d^{3}\,x$.
The choice of boxes is the same for all the graphs considered.
Each box contains a large number of vertices of the semiclassical
state (${\cal L}>\!>\ell _{P}$), but it is considered as
infinitesimal in the scale where the space can be regarded as
continuous. The sum over the vertices in (\ref{REGHAM}) is
subsequently split as a sum over the vertices in each box, plus a
sum over the boxes. Also, one assumes that the gauge or matter
operators are slowly varying within a   box (${\cal L}<\!<\lambda
$, with $\lambda$ been the correspondent particle wavelength), in
such a way that for all the vertices inside a given  box one can
write
$\langle \Gamma, S|\dots \underline{{%
\hat{F}}}_{ab}({\vec V})\dots |\Gamma, S \rangle = \mu {\underline
F}_{ab}({\vec{x}})$, for example. Here ${\ {\underline F}}_{ab}$
is the classical electromagnetic field at the center of the box
and $\mu$ is a dimensionless constant which is determined in such
a way that the standard classical result in the zeroth order
approximation is recovered. Applying  the procedure just described
to (\ref{REGHAM}) leads to {
\begin{eqnarray}{
{H}_\Gamma^{B}}&=&{ \sum_{{\rm Box}}}\,\,\left( \frac{}{}
\partial^{a_1}\partial^{a_2}\dots\partial^{a_k}\dots{ {\underline
F}}_{p_1q_1}({\vec x})\right)\,\dots \left(\dots \, \frac{}{}
\partial^{a_{m-1}}
\partial^{a_m}{{
{\underline F}}_{pq}({\vec x})}\right) \times \nonumber \\
&& \sum_{{\vec V}\in {\rm Box}} {\ell _{P}}^{{3}}\sum_{{{\vec
V}(\Delta )={\vec V}}}\,\mu^{{n+1}} {{\langle \Gamma,S}|
\frac{1}{\ell _{{ P}}^{3}}{{\hat{T} }_{a_{1}\dots a_{m}
}{}^{\,pq\dots p_{1}q_{1} }({\vec V},{\vec U}(\Delta ), A_{ia}
)|\Gamma, S \rangle},}\nonumber \\ \label{HAMG}
\end{eqnarray}}
where $n+1$ is the total number of factors $F_{rs}({\vec x})$,
each of which can contain some derivatives. The expectation value
of the gravitational contribution is supposed to be a rapidly
varying function inside each box. Next we consider the total
effective Hamiltonian (\ref{TEFFHAM}), which is defined as an
average over the graphs $\Gamma$, i.e. over the adapted
triangulations. This effectively amounts to average the remaining
expectation values in each box of the sum
(\ref{HAMG}). We call this box-average ${{t}%
}_{a_{1}\dots a_m }{}^{\,pq\dots p_{1}q_{1} }({\vec x})$ and
define it  by \ba \sum_{\Gamma}P(\Gamma) \sum_{{\vec V}\in {\rm
Box}} {\ell _{P}}^{{3}}\sum_{{{\vec V}(\Delta )={\vec V}}}
{{\langle \Gamma,S}| \frac{1}{\ell _{{ P}}^{3}}{{\hat{T}
}_{a_{1}\dots a_{m} }{}^{\,pq\dots p_{1}\,q_{1} }|\Gamma, S
\rangle}}=\nonumber \\
d^3x  \, {t}_{a_{1}\dots a_m }{}^{\,pq\dots p_{1}q_{1} }({\vec
x}).\ea We estimate ${t}_{a_{1}\dots a_m }{}^{\,pq\dots p_{1}q_{1}
}({\vec x})$  by demanding it to be constructed from the flat
space tensors $\delta_{ab}$ and $ \epsilon_{abc}$, together with
the corresponding ones $\tau_i$ and $\delta^a_i,$ in the case of
spinors. In this way we are imposing isotropy and rotational
invariance on our final Hamiltonian in the frame selected by the
choice of the semiclassical states. This is somewhat analogous to
the spontaneously symmetry breaking scheme, where the choice of
the vacuum selects a particular frame for the symmetry breaking
vacuum expectation values. Also, the scalings given in
(\ref{EXPV}) together with the additional assumptions \be\langle
\Gamma, S|...{\hat V}...|\Gamma, S\rangle \longrightarrow
\ell_P^3, \qquad U_I^a \longrightarrow \ell_P, \ee
 are used in the above  estimation. After
replacing the summation over boxes by the integral over space, the
resulting Hamiltonian has the final form
\ba{H}^{B}=&&{\int d^{3}x
\ \left( \frac{}{}
\partial^{a_1}\partial^{a_2}\dots \partial^{a_k}\dots{ {\underline
F}}_{p_1q_1}({\vec x})\right)\dots \,\,\left(\frac{}{}\dots
\partial^{a_{m-1}}
\partial^{a_m}{{
{\underline F}}_{pq}({\vec x})}\right)}\nonumber \\
&&\qquad \times   {t}_{a_{1}\dots a_{m} }{}^{\,pq \dots
p_{1}\,q_{1} }({\vec{x}}). \ea

Finally we exhibit some of the required  expansions. The basic
electromagnetic holonomy around the path $\alpha _{JK}$ \be
{\underline{h}}_{\alpha _{JK}(\Delta
)}=e^{-i\int_{\alpha_{JK}(\Delta )}dt\,
{\dot{U}^{a}(t)}\underline{\hat{A}}_{a}({s(t)})}=e^{-i{
{\Phi}}^{B}({F_{JK}})}, \ee can be written in terms of the
magnetic flux ${\Phi}^{B}$ through the face bounded by such closed
path. The expansion of such flux with respect to the magnetic
field at the vertex is
\begin{eqnarray}
&&{\Phi ^{B}(F_{IJ}) =\left(
1\,\,+\frac{1}{3}{(U_{I}^{c}+U_{J}^{c})}\,\partial
_{c}\,+\frac{1}{12}{(U_{I}^{c}\,U_{I}^{d}+U_{I}^{c}\,U_{J}^{d}+U_{J}^{c}
\,U_{J}^{d})}\right.} \nonumber \\
&&{\left. \qquad \qquad \qquad \times \,\,\partial _{c}\,
\partial _{d}+\frac{}{}...\right)  \,
\frac{1}{2}{U_{I}^{a}U_{J}^{b}}\epsilon _{abc}B^{c}({\vec V})},
\nonumber
\end{eqnarray}
to second order in the derivatives.

The expansion of the gravitational operator (\ref{GRAVOP}) is \be
{\hat w}_{i \, L \Delta }= {U^a_L} w_{ia} + {U^a_L U^b_L} w_{iab}
+ {U^a_L U^b_L U^c_L} w_{iabc} + {\cal O}(s^4w), \ee to third
order in the vectors $U^a_L$. The remaining operators  in the
expansion include the connection together with the volume
operators, in the form
\begin{eqnarray}
{w_{ia}} &=& {\frac{1}{2} [A_{ia},\sqrt{V}],\quad w_{iab} =
\frac{1}{8}
\epsilon_{ijk} [A_{ja},[A_{kb},\sqrt{V}]],}  \nonumber \\
&& {w_{iabc} = -\frac{1}{48} [A_{ja},[A_{jb},[A_{ic},\sqrt{V}]]]}.
\end{eqnarray}
\subsection{The results}
The application of the method described in the previous section
leads to the following effective Hamiltonians. In the
electromagnetic case we obtain \cite{AMU1}
\begin{eqnarray} {{H}^{EM}}&=& {\frac{1}{Q^2}\int
d^3{\vec x}} \left[\left(1+ {\theta_7 }\,{\left(\frac{\ell_{
P}}{{\cal L}}\right)^{{2+2 \Upsilon}}}
\right)\frac{1}{2}\left(\frac{}{}
{\underline{{\vec B}}^2 + \underline{{\vec E}}^2}\right) \right.  \nonumber \\
&&\left. + {{\theta_8 {\ell_P}}} \left( \frac{}{} {\underline{\vec B}%
\cdot(\vec\nabla \times\underline{\vec B})+ \underline{\vec E}%
\cdot(\vec\nabla \times\underline{\vec E}) }\right)\right.  \nonumber \\
&&\left.+ {{\theta_3 \, {\ell_P^2}} }\, \left({ \frac{}{}\underline{B}%
^a \,\nabla^2 \underline{B}_a + \underline{E}^a \,\nabla^2 \underline{E}%
_a}\right)
+ {{\theta_2\,{\ell_P^2}}}\,{{\underline E}^a \partial_a \partial_b {%
\underline E}^b} \ \right.  \nonumber \\
&&\left. + {{\theta_4\,  \ell_P^2 {\cal L}^2 }\, \,{ \left(
\frac{\ell_P}{{\cal L}}\right)}^{{2\Upsilon}}
}\, \, {\left(\frac{}{}%
\underline{{\vec B}}^2\right)^2} +\dots \right]. \label{TOTALMAX}
\end{eqnarray}
The above  result has been extended to the Yang-Mills case in
Ref.\cite{MARAT}. The case where  only $\theta_8$ in non zero
corresponds to the Gambini-Pullin effective Hamiltonian
\cite{GPED}.

In the case of Majorana fermions with mass $m$ we have
\cite{AMU0,AMU2}
\begin{eqnarray}
{{ H}^{F}}= {\int d^3 x }\left[ {i \ \pi(\vec x)}
 \  \left({1 + {{ \kappa}_{1}
\left(\frac{\ell_P}{{\cal L}}\right)^{1+\Upsilon}} +
{\kappa_2\left(\frac{\ell_P}{{\cal L}}\right)^{2+2\Upsilon}}+}
\right. \right.&& \nonumber\\
 \left. \left. {{+ \frac{\kappa_3}{2} \
{\ell_P^2 }\ \nabla^2 }} \right)\, {\tau^d\partial_d \, \xi({\vec x})}\right. && \nonumber \\
+ \left.{{\frac{i}{4} {\frac{1}{{\cal L}}}}\ \pi({\vec x})}
\left({{\kappa_4 {\left(\frac{\ell_P}{{\cal L}}\right)^\Upsilon }
+
 \kappa_5 {\left(\frac{\ell_P}{{\cal L}}\right)^{1+ 2\Upsilon}}+ \dots
 + \frac{\kappa_{7}}{2} \,
{\ell_P^2}
\nabla^2}} \right){\xi({\vec x})}\right. && \nonumber\\
\left. + {\frac{m}{2 } \xi^T({\vec x})\ (i \sigma^2)} {\left(1 +
{{\kappa} _{8}{\left(\frac{\ell_P}{{\cal L}}\right)^{1+\Upsilon}}}
+ {\left(\frac{}{} \kappa_{9}{\ell_P}+\dots \right)}{ \tau^a
\partial_a }\right)\xi({\vec x})} + c.c.
\frac{}{}\right],\nonumber \\ \label{TOTALWEYL}
\end{eqnarray}
where $\pi(\vec x)= i \xi^*(\vec x)$ is the canonically conjugated
momentum of the two-component spin 1/2 field $\xi(\vec x)$. The
notation is ${\tau}^a= -(i/2) {\sigma^a}$, where ${ \sigma}^a$ are
the standard Pauli matrices.

Some comments regarding the procedure are now in order. (i) The
dimensionless numerical coefficients $\theta_A$, $\kappa_A$
appearing in the above effective Hamiltonians remain arbitrary in
the procedure, but are independent of any parameter of the model.
To predict them will require an exact knowledge of the states used
to calculate the corresponding average values, together with the
action of the basic operators upon them. In this respect  we have
obtained  a systematic parameterization of the possible modified
Hamiltonians in terms of higher derivative operators, where the
dependence upon the scales of the problem ($\ell_P, {\cal L},
\Upsilon $) has been explicitly determined. In particular, these
coefficients could turn out to be zero, leading to no dynamical
corrections. (ii) A main drawback of the method is that it does
not incorporate properly the dynamics via the Hamiltonian
constraint of quantum gravity. An improved semiclassical
approximation taking care of the many issues assumed or left over
in our approach is certainly needed. (iii) There is also the
question of interpreting the results (\ref{TOTALMAX}) and
(\ref{TOTALWEYL}) in relation to their transformation properties
under standard active (particle)global Lorentz transformations.
There are various possibilities. These Hamiltonians have been
calculated in a specific reference frame where isotropy under
rotations is maintained. This could signal the presence of an
absolute reference frame, a reborn version of the ether. This
hypothesis will be further explored in the following and has
received a lot of attention in the framework of different models
in the literature. On the other hand one could argue that this
specific frame has been selected by the choice of the
semiclassical states that describe the flat space limit, in
analogous way to the spontaneously symmetry breaking procedure.
The transformation properties under this assumption have not been
studied. The third alternative is that only  a complete
calculation of the corrections to flat space dynamics will tell us
which is the proper generalization, if any, of the global Lorentz
transformations, thus yielding a deformed or extended version for
preserving the equivalence of inertial frames and leading to a
modified relativity principle. (iv) Finally we must say that the
above results have provided some motivation for phenomenological
theories which have explored the consequences of such
modifications and have shown that either experimental or
astrophysical observations, even with the actual level of
sensitivity, can set rather stringent bounds upon the correction
parameters. In other words, Planck scale sensitivity has been
already attained in a rather broad set of observations.
Conversely,  this would imply severe constraints upon the
fundamental theory once the semiclassical limit is correctly
performed.
\subsection{The parameters $\cal L$ and $\Upsilon$}
\label{subsec:param}
In order to produce numerical estimations for
some of the effects arising from the previously obtained
modifications to flat space dynamics, we must further fix the
value of the scales $ {\mathcal L}$ and $\Upsilon$.

Let us recall that ${\mathcal L}$ is a scale indicating the onset
of the distance from where the non perturbative states in the loop
representation can be approximated by the classical flat metric.
The propagating particle  is characterized by energies which probe
distances of the order of the De Broglie wave length $\lambda$. As
mentioned previously,  just to be consistent with a description in
terms of classical continuous equations it is necessary to require
that ${ {\mathcal L}}< \lambda$. Two distinguished cases for
choosing $\cal L$ arise: (i) the mobile scale, where we take the
marginal choice ${ {\mathcal L}}= \lambda$ in each situation and
(ii) the universal scale, which has been introduced in the
discussion of the GZK anomaly \cite{ALFARO}. The consideration of
the different reactions involved produces the  preferred range $
4.6\times 10^{-8}\, GeV^{-1}\geq{ {\mathcal L}}\geq 8.3 \times
10^{-9}\, GeV^{-1}$. A recent study of the gravitational Cerenkov
effect together with neutrino oscillations \cite{LAMBIASE} yields
a universal scale evaluation which is consistent with the former.

Ranges for $\Upsilon$ have been estimated considering  the
observation that atmospheric neutrino oscillations at average
energies of the order $10^{-2}-10^2 $ GeV are dominated by the
corresponding mass differences via the oscillation length $L_m$
\cite{BRUSTEIN}.
 This means that additional contributions to the oscillation
length, in particular the quantum gravity correction $L_{QG}$,
should satisfy $L_{QG}> L_m$. This has been used to set  lower
bounds upon $\Upsilon$.  Within the proposed two  methods for
estimating the scale ${{\mathcal L}}$ of the process we obtain:
(i) $\Upsilon
> 0.15$ when ${ {\mathcal L}}$ is considered as a mobile scale and
(ii) $\Upsilon > 1.2  $ when the scale ${ {\mathcal L}}$ takes the
universal value ${{\mathcal L}}\approx 10^{-8}\, GeV^{-1}$
\cite{AMU2}. These results are summarized in Table \ref{tab:1}.
\begin{table}
\centering \caption{Ranges for the parameters ${\cal L}$ and
$\Upsilon$} \label{tab:1}
\begin{tabular}{lll}
\hline\noalign{\smallskip}
 & $\qquad \quad \quad {\cal L}\, [GeV^{-1}]$ & $\qquad \quad {\Upsilon}$ \\
\noalign{\smallskip}\hline\noalign{\smallskip}
Mobile scale $\qquad$ & $\quad \qquad \qquad {\cal L}= \lambda$ &$\qquad {\Upsilon}\,>\, 0.15$  \\
Universal scale $\qquad$ & $8.3\times 10^{-9}\,<\,{\cal L} \,<\,4.6\times 10^{-8}$ & $\qquad {\Upsilon}\,>\, 1.2 $\\
\noalign{\smallskip}\hline
\end{tabular}
\end{table}
\section{Phenomenological aspects}
\label{section:4} In this section we discuss four phenomenological
applications of models presenting Planck scale dynamical
corrections.  In subsection \ref{ssec:bounds}  we summarize the
derivation of stringent bounds upon combinations of some
parameters appearing in the effective fermionic Hamiltonian
(\ref{TOTALWEYL}), using existing data from clock-comparison
experiments.
 Subsection \ref{ssec:radcorr} is devoted to the calculation
of radiative corrections arising in  LIV effects with preferred
frames associated to space granularity. In order to make the
calculation simple we consider a Yukawa model for the
interactions. Subsection \ref{ssec:radliv} contains the discussion
of synchrotron radiation in Planck scale modified electrodynamics,
emphasizing the model of Ref. \cite{MP} which is
phenomenologically formulated as a theory parameterizing LIV with
dimension five operators. Finally, a unified description of
radiation in Planck scale modified electrodynamics incorporating
different models is presented in subsection \ref{ss102}.
\subsection{Bounds on the fermionic sector parameters from clock-comparison
experiments} \label{ssec:bounds} The Hamiltonian (\ref{TOTALWEYL})
for two components fermions was obtained under the assumption of
flat space isotropy and it was assumed to account for the dynamics
in a preferred reference frame, identified as the one in which the
Cosmic Microwave Background looks isotropic. The earth velocity
$\mathbf w$ with respect to that frame has already been determined
to be $w/c\approx 1.23 \times 10^{-3}$ by COBE \cite{COBE}. Thus,
in the earth reference frame one expects the appearance of signals
indicating minute violations of space isotropy encoded in the
diurnal variation of the $\mathbf w$-dependent terms appearing in
the  transformed Hamiltonian or Lagrangian \cite{SUV}.  On the
other hand, many high precision experimental test of such
variations, using atomic and nuclear systems for example, have
been already reported in the literature
\cite{HUDREV,EXP1,EXP2,EXP3} and already analyzed in terms of the
SME. Amazingly enough such precision is already adequate to set
very stringent bounds on some of the parameters arising from the
quantum gravity corrections.

We have considered the case of non-relativistic Dirac particles
and obtained corrections which involve the coupling of the spin to
the CMB velocity, together with a quadrupolar anisotropy of the
inertial mass \cite{SUV}. The calculation was made with the
choices $\Upsilon=0$ and ${ {\mathcal L}}=1/M$, where $M$ is the
rest mass of the fermion. Here it is important to emphasize again
that $\mathcal L \, >> \, \ell_P$ is not the scale of quantum
gravity but rather the lower limit of distances from which  a
description of space as a continuum is already valid. In this
sense we are taking the upper limit settled by the de Broglie wave
length $\lambda_D=1/M$ of the nucleon, which allows us to make
sense of the corresponding Dirac equation. As we will show later,
a lower limit for ${\mathcal L}$ will only make the bounds upon
the LIV parameters more stringent.

Keeping only terms linear in $\ell_P$, the equations of motion for
a  Majorana fermion of mass $m$ described by the two-component
spinor $\xi$, arising from the Hamiltonian (\ref{TOTALWEYL}) are
given by \ba &&\left[i\partial/\partial t -iA {\vec
\sigma}\cdot{\vec \nabla }+K/2 \right]\xi -m(\alpha -i\beta {\vec
\sigma}\cdot{\vec \nabla } )\chi=0, \nonumber \\
&& \left[i\partial/\partial t + iA {\vec \sigma}\cdot{\vec \nabla
}-K/2 \right]\chi -m(\alpha -i\beta {\vec \sigma}\cdot{\vec
\nabla} )\xi=0,\label{MAJF}\ea where \ba &&A=(1+\Theta_1 m
\ell_p), \quad
\alpha=(1+\Theta_3 m \ell_P),\nonumber \\
&& K=m \Theta_4 m \ell_P, \quad \beta= \Theta_2 \ell_P. \ea  The
notation in Eqs. (\ref{MAJF}) is $\chi=-i\sigma_2\,\xi^*$ and it
is a direct matter to verify the consistency between them. These
equations can be readily extended to the Dirac case by considering
$\chi$ and $\xi$ as independent spinors unified in
$\Psi^T=(\xi^T,\, \chi^T)$, with the result
\begin{equation}
\left( i\gamma ^{\mu }\partial _{\mu }\,+\Theta_1 m\ell
_{P}\;i\mathbf{\gamma}\cdot \nabla -\frac{K}{2}\gamma _{5}\gamma
^{0}-m\left( \alpha -i\Theta_2 \ell _{P}\,{{\Sigma}}\cdot \nabla
\right) \right) \Psi =0. \label{DIREQ0}
\end{equation}
Here we have used the representation in which $\gamma _{5}$ is
diagonal and the spin operator is\ $\Sigma ^{k}=(i/2)\epsilon
_{klm}\gamma ^{l}\gamma ^{m}$, with standard particle physicist
conventions. The normalization has been chosen so that in the
limit $(m\ell_P)\rightarrow 0$ we recover the standard massive
Dirac equation. The term $m\left( 1+{\Theta_3\;}m\ell _{P}\right)
$ can be interpreted as a renormalization of the mass whose
physical value is taken to be $M=m\left( 1+{\Theta_3\;}m\ell
_{P}\right) $. After these modifications we can write an effective
Lagrangian describing the time evolution as seen in the CMB frame.
In order to obtain the dynamics  in the laboratory frame we
implement an observer (passive) Lorentz transformation in the
former Lagrangian and rewrite it  in a covariant looking form by
introducing explicitly the CMB frame four velocity $W^{\mu
}=\gamma (1,\,{\mathbf{w}}/c)$ arising from the boosted  rest
frame velocity $(1, {\mathbf 0})$. The result is
\begin{eqnarray}
L_{D}&=&\frac{1}{2}i\bar{\Psi}\gamma ^{\mu }\partial _{\mu }\Psi -\frac{1}{2}M%
\bar{\Psi}\,\Psi +\frac{1}{2}i(\Theta_1 M\ell
_{P})\bar{\Psi}\gamma _{\mu }\left(
g^{\mu \nu }-W^{\mu }W^{\nu }\right) \partial _{\nu }\Psi \nonumber \\
&&+\frac{1}{4}(\Theta_2M\ell_{P})\bar{\Psi}\epsilon _{\mu \nu
\alpha \beta }W ^{\mu }\gamma ^{\nu }\gamma ^{\alpha }\partial
^{\beta }\Psi - \frac{1}{4}(\Theta_4 M\ell_P)M W_\mu
\bar{\Psi}\gamma_5\gamma^\mu\Psi +h.c. ,\nonumber \\ \label{otra}
\end{eqnarray}
where $W^\mu$ is an external  non-dynamical field which is not to
be varied in the corresponding action. It is interesting to remark
that the above Lagrangian provides a specific realization of the
general form considered in the SME. According to Ref.
\cite{FERMSME}, the identifications are \ba
&&a_\mu=H_{\mu\nu}=d_{\mu\nu}=e_\mu=f_\mu=0, \qquad
c_{\mu\nu}=\Theta_1 M \ell_P(g_{\mu\nu}-W_\mu W_\nu), \label{ID}
\nonumber\\
&&g_{\alpha\beta\gamma}=-\Theta_2 M \ell_P W^\rho
\epsilon_{\rho\alpha\beta\gamma}, \qquad b_\mu=
\frac{1}{2}\Theta_4 M^2 \ell_P W_\mu. \ea The above reference
provides also the non-relativistic limit of the Hamiltonian
corresponding to the general SME modified fermion Lagrangian. In
our case, up to first order in $\ell_P$ and up to order ${({\bf
w})/c}^{2}$, the identifications (\ref{ID}) yield
\begin{eqnarray}
\tilde{H}=  Mc^2(1+\Theta_1\, M\ell_P\,\left({\mathbf{w}}/{c}
\right) ^{2})+ \Theta_1M\ell _{P} %
\left[ \frac{{\bf w}\cdot {Q}_{P}\cdot {\bf w}}{Mc^2}\right] &&\nonumber\\
+ \left(1+2\,\Theta_1M\ell
_{P}\left(1+\frac{5}{6}\left({\mathbf{w}}/{c} \right)
^{2}\right)\right)\left(\frac{p^{2}}{2M}+g\,\mu
\,{\bf s}%
\cdot {\bf B}\right)&&\nonumber\\
 + \left(\Theta_2+\frac{1}{2}\Theta_4 \right)M\ell
_{P}\left[\left(2Mc^2
-\frac{2p^{2}}{3M}\right)\,{\bf s}%
\cdot \frac{\bf w}{c}+\frac{1}{M}\,{\bf s}\cdot {Q}_{P}\cdot
\frac{\bf w}{c}\right] ,&& \label{B6}
\end{eqnarray}
where ${ s^i}={ \sigma^i}/2$.

The above effective Hamiltonian has been employed in the
description of the valence nucleon responsible for the transitions
measured in clock-comparison experiments using pairs of nuclei
like $({}^{21}Ne,{}^3He)$ \cite{EXP2}, and $({}^{129}Xe,{}^3He) $
\cite{EXP3}, for example. In (\ref{B6}) we have not written the
terms linear in the momentum since they average to zero in the
nuclear bound state situation. Here $g$ is the standard
gyromagnetic factor, and $Q_{P}$ is the momentum quadrupole tensor
with components $Q_{Pij}=p_{i}p_{j}-1/3p^{2}\delta _{ij}$. The
terms in the second square bracket of the LHS of (\ref{B6})
represent a coupling of the spin to the velocity of the privileged
reference frame. The first term inside the bracket has been
measured with high accuracy and an upper bound for the coefficient
has been found. The second term in the same bracket is a small
anisotropy contribution and can be neglected. Thus we find the
correction \be \delta H_{S}= \left(\Theta_2
+\frac{1}{2}\Theta_4\right) M\ell _{P} (2Mc^2) \left[ 1 + O\left(
\frac{p^{2}}{2M^2c^2}\right) \right] \mathbf{s}\cdot
\frac{\mathbf{w}}{c}. \label{DHS}\ee

The first square bracket in the LHS of (\ref{B6})  represents an
anisotropy of the inertial mass and has been bounded in
Hughes-Drever like experiments. With the approximation
$Q_{P}=-5/3<p^{2}>Q/R^{2}$ for the momentum quadrupole moment,
with $Q$ being the electric quadrupole moment and $R$ the nuclear
radius, we obtain
\begin{equation}
\delta H_{Q}=-\Theta_1 M\ell _{P}\frac{5}{3}\left\langle \frac{p^{2}}{2M}%
\right\rangle \left( \frac{Q}{R^{2}}\right) \left(
\frac{w}{c}\right) ^{2}P_{2}(\cos \theta ), \label{QMM}
\end{equation}
for the quadrupole mass perturbation, where $\theta$ is the angle
between the quantization axis and $\mathbf{w}$. Using
$<p^{2}/2M>\sim 40$ MeV for the energy of a nucleon in the outer
shell of a typical heavy nucleus, together with the experimental
bounds of references \cite{EXP2,EXP3} we find \cite{SUV}
\begin{equation}
\mid \Theta_2+\frac{1}{2}\Theta_4\mid <2\times 10^{-9},\qquad \mid
\Theta_1\mid <3\times 10^{-5}. \label{Result2}
\end{equation}
From Eqs. (\ref{DHS}) and (\ref{QMM}) we realize that only
products of the type (LIV pa\-ram\-e\-ters)$\, \times\,
1/\mathcal{L}\,\approx {\rm (LIV pa\-ram\-e\-ters)}\times M$ are
bounded by the experiments. In this way choosing a smaller
$\mathcal{L}$, i.e. a larger $M$, will just decrease the upper
bound for the LIV parameters. Thus the choice
$\mathcal{L}=\lambda_D=1/M $ produces the weakest bound in this
analysis.
\subsection{Radiative corrections in preferred frames modelling
space granularity} \label{ssec:radcorr} So far  corrections to the
dynamics have arisen only in the non-interacting theory, through
factors of the type {$(E/M_P)^\Upsilon$},  which are directly
relevant at unaccessible energies $E \approx M_P$. A possibility
of probing such high energies is through the inclusion of
{radiative corrections (particle's self energies, for example )}
because the internal momenta are integrated up to the maximum
allowed in a given reference frame. The standard folklore with
respect to any new physics entering at high scales (here Planck
scale) is that it has negligible effects on the leading-order
low-energy physics (here free particle corrections). Contrary to
this belief, we  show in this subsection  that modelling space
granularity via the introduction of a physical cutoff, which
defines a preferred reference frame, leads to unsuppressed
dimension four LIV contributions \cite{CPSUV,CPSUVUN}. Results
consistent with these have been obtained in Ref. \cite{MP}.

To this end we consider the calculation of one-loop radiative
corrections in the Yukawa model \be L = 1/2 \, (\partial\phi)^2-
1/2\, \mu^2 \phi^2 + \bar\psi (i\gamma^\mu\partial_\mu - m) \psi +
g \phi\bar\psi\psi + (\rm{LVT}). \ee Here LVT refers to the highly
suppressed zeroth order Lorentz violating terms that take into
account the previously discussed free particle dynamical
modifications. We model the space granularity by introducing a
physical cut-off $\Lambda$ in such a way that the magnitude of the
three-momentum in any loop is bounded by this quantity. The
parameter $\Lambda$ defines the onset of the scale at which the
granularity of space becomes manifest.  A convenient way of
incorporating this requirement is to introduce the physical cutoff
function \be f( |{\vec k}|/\Lambda), \quad f(0)=1, \quad
f(\infty)=0,\quad \Lambda\approx 1/{\cal L}, \label{CUTOFFF}\ee
which suppresses the internal momenta having $|{\vec k}| \geq
\Lambda$. One can visualize this function as a smoothed theta
function with the correspondent  jump at $|{\vec k}|\approx
\Lambda$.  Our choice is that the cutoff function depends on
spatial momentum, but not of energy, and that it is rotationally
invariant.
\paragraph{The fermion self energy \cite{CPSUVUN}}
Tests of Lorentz invariance typically deal with the relation
between energy and 3-momentum of an isolated particle, which is
given by the position of the pole of the particle's full
propagator.  We will calculate the effect of loop corrections on
this relation. It is convenient, as usual, to write the full
fermion propagator as
\begin{equation}
   S(p) = \frac{i}{p^{\mu}\gamma_{\mu} - m - \Sigma_2(p)},
\end{equation}
where $\Sigma_2(p)$ is the standard fermion self energy. Here we
are neglecting the zeroth order corrections to the dispersion
relations for the free particles.

As a first manner to model  the validity of a continuous
description of space only up to some short distance $1/\Lambda
\approx {\cal L} $, we choose to cutoff only the free scalar
propagator, but not the free fermion propagator, which is
sufficient to cutoff the UV divergence of the one-loop fermion
self-energy while giving a maximally simple calculation. In this
way, the one-loop approximation to the fermion  self-energy is
\begin{equation}
 \Sigma_2(p) =
   ig^2\int
   \frac{ {\rm d}^4k }
        {(2\pi)^4}
   \, \frac{ f(|\vec{k}|/\Lambda) }
        { k^2+i\epsilon }
   \, \frac{\gamma_{\rho}(p+k)^{\rho}+ m}
        { (p+k)^2-m^2+i\epsilon },
  \end{equation}
where we have set $\mu=0$ for simplicity. We wish to investigate
the self-energy when the momentum $p^\mu$ and the mass $m$ are
much less than the parameter $\Lambda$.  If the integral were
convergent as the
{parameter}
is removed ($\Lambda\to\infty$), then the self-energy would equal
its value at $\Lambda=\infty$ plus corrections of order
$m/\Lambda$ and $p/\Lambda$.  However, it is easy to see that the
integral is divergent in this limit.  To analyze its behavior, we
observe that differentiating it twice with respect to $p$ gives a
finite integral (when $\Lambda\to\infty$).  So we extract the
zeroth order and the linear terms in the  momentum expansion of
$\Sigma_2$ about $p=0$
\begin{eqnarray}
\label{eq:Sigma.exp} \Sigma_2(p) &=& \Sigma_2(p=0)
          + p^0 \partial_{p^0}\Sigma_2(p)|_{p=0}
          + p^i \partial_{p^i}\Sigma_2(p)|_{p=0}
          + \hat\Sigma(p).
\end{eqnarray}
Now $\hat\Sigma(p)$ is $O(p^2)$ as $p\to0$ and it is finite when
the cutoff is removed.  Therefore it only gives power-suppressed
Lorentz violations.  The unsuppressed Lorentz violations, if any,
can only arise from  the first three terms. We find that the terms
$\Sigma_2(p=0)$, $\partial_{p^0}\Sigma_2(p)|_{p=0}$ and
$\partial_{p^i}\Sigma_2(p)|_{p=0}$ are proportional to the Dirac
matrices $1$, $\gamma_i$, $\gamma^0$, exactly as in the
Lorentz-invariant case; this follows from the discrete symmetries
of our model, rather than full Lorentz invariance.  Hence we can
write
\begin{eqnarray}
\label{eq:Sigma.exp1} \Sigma_2(p)&=& A m + p^0\gamma_0 B +
p^i\gamma_i C + \hat\Sigma(p, \Lambda=\infty) + O(p/\Lambda,
m/\Lambda)
\nonumber\\
      &=& A m + p^\mu\gamma_\mu C + p^0\gamma_0 (B-C) + \hat\Sigma(p, \Lambda=\infty)
 + O(p/\Lambda, m/\Lambda),\nonumber \\
\end{eqnarray}
where $A$, $B$ and $C$ are numerical-valued functions of the
parameters of the problem.  The only leading-power Lorentz
violation is caused by the difference between the $B$ and $C$,
which we calculate below and find it to be nonzero and
unsuppressed.

In fact, the constant term is
\begin{equation}
 \Sigma_2(0) =
   \frac{ig^2}{(2\pi)^4} \int {\rm d}^4k \,
   f(|\vec{k}|/\Lambda)\, \frac{  \gamma_{\rho}k^\rho+ m }
        { (k^2-m^2+i\epsilon) \, (k^2+i\epsilon) }.
\end{equation}
Since the physical cutoff depends only on $|{\vec k}|$, the
modified theory is invariant under reversal of any component of
$k^\mu$ and the term proportional to $\gamma_\rho$ vanishes after
integration. Hence
\begin{eqnarray}
  \label{eq:A}
  A &=&
   \frac{ig^2}{(2\pi)^4} \int {\rm d}^4k \,
   \frac{ f(|\vec{k}|/\Lambda) }
        { (k^2-m^2+i\epsilon) \, (k^2+i\epsilon) }.
\nonumber\\
   &=&
   \frac{g^2}{4\pi^2} \int_0^\infty {\rm d}l \, f(l/\Lambda) \,
   \frac{l^2}{m^2}\left( \frac{1}{l} - \frac{1}{E_l} \right) ,
\end{eqnarray}
where $E_l=\sqrt{l^2+m^2}$. In the second line, we have performed
the integrals over $k^0$ and the angle of $\vec{k}$.  The
corresponding term in $\Sigma_2$ is proportional to the unit Dirac
matrix, so that it is equivalent to a fermion mass term. Note that
when $\Lambda\to\infty$ there is a logarithmic divergence, which
can be removed by the usual mass renormalization.  Since the term
is proportional to the unit Dirac matrix, it does not violate
Lorentz invariance.

The situation for the derivative terms is different.  For the
{time-like}
derivative we have
\begin{eqnarray}
\partial_{p^0}\Sigma_2(p)|_{p=0} &=&
   \frac{ig^2}{(2\pi)^4} \int {\rm d}^4k \,
   \frac{ f(|\vec{k}|/\Lambda) }{ k^2+i\epsilon }
   \left[
        \frac{ \gamma_0 }{ k^2-m^2+i\epsilon }
        - \frac{ 2k^0 (\gamma_\rho k^\rho+ m) }{ (k^2-m^2+i\epsilon)^2
        }
   \right].\nonumber \\
\end{eqnarray}
Invariance under reversal of components removes all but the term
proportional to $\gamma_0$, so that
\begin{eqnarray}
\label{eq:B}
 B &=&
   \frac{ig^2}{(2\pi)^4} \int {\rm d}^4k \,
   \frac{ f(|\vec{k}|/\Lambda) }{ k^2+i\epsilon }
   \left[
        \frac{ 1 }{ k^2-m^2+i\epsilon }
        - \frac{ 2{k^0}^2 }{ (k^2-m^2+i\epsilon)^2 }
   \right]
\nonumber\\
  &=&
   \frac{g^2}{4\pi^2} \int_0^\infty {\rm d}l \, f(l/\Lambda)
   \left[
       \frac{2l^2(E_l-l)}{m^4} - \frac{l}{m^2}
   \right]
.
\end{eqnarray}
Similarly, from the
{space-like}
derivative we get
\begin{eqnarray}
\label{eq:C}
 C &=&
   \frac{ig^2}{(2\pi)^4} \int {\rm d}^4k \,
   \frac{ f(|\vec{k}|/\Lambda) }{ k^2+i\epsilon }
   \left[
        \frac{ 1 }{ k^2-m^2+i\epsilon }
        + \frac{ 2\vec{k}^2/3 }{ (k^2-m^2+i\epsilon)^2 }
   \right]
\nonumber\\
  &=&
   \frac{g^2}{4\pi^2} \int_0^\infty {\rm d}l \, f(l/\Lambda)
   \left[
       \frac{2l^3}{3m^4}
       -\frac{l}{m^2}
       +\frac{l^2}{m^2E_l}
       -\frac{2l^4}{3m^4E_l}
       -\frac{l^4}{3m^2E_l^3}
   \right]
.
\end{eqnarray}
As expected, both $B$ and $C$ have logarithmic divergences as
$\Lambda\to\infty$.

We are interested in the difference, which gives the low-energy
violation of Lorentz invariance
\begin{eqnarray}
\label{eq:B-C} B-C &=&
   - \frac{2ig^2}{(2\pi)^4} \int {\rm d}^4k \,
   \frac{ f(|\vec{k}|/\Lambda) }{ k^2+i\epsilon }
   \, \frac{ \tfrac13 \vec{k}^2 + {k^0}^2}{ (k^2-m^2+i\epsilon)^2 }
\label{int1}\\
  &=&
   \frac{g^2}{12\pi^2} \int_0^\infty {\rm d}l \, f(l/\Lambda)
   \left[
       \frac{8E_l^3}{m^4}
       -\frac{8l^3}{m^4}
       -\frac{12E_l}{m^2}
       +\frac{3}{E_l}
       +\frac{m^2}{E_l^3}
   \right]
. \label{int2}
\end{eqnarray}
Suppose the integral (\ref{int1}) were convergent in the continuum
limit. Then it would be zero in this limit, as can be seen by Wick
rotation of an integral of $\frac13 \vec{k}^2 + {k^0}^2$ times an
arbitrary Lorentz-invariant function $F(k^2)$. This argument
depends on the integral being absolutely convergent, i.e., on the
integral of the absolute value of the integrand being finite. In
contrast, the corresponding integral for $B-C$ is logarithmically
divergent.  So the leading term for the integral, obtained when
$\Lambda\to\infty$, depends on how this  cutoff is removed. In our
case this is dictated by the physical modelling of the space
granularity via the function (\ref{CUTOFFF}), as expressed in
Eq.(\ref{int2}). This choice implies that  the factor in square
brackets behaves like $m^2/l^3$ when $l/m\to\infty$, so that the
integral is finite when the cutoff is removed. Besides, we can
write this square bracket as  indicated in the second line of the
following equation, which leads to a direct calculation of the
leading term
\begin{eqnarray}
  \label{eq:B-C.limit}
  \lim_{\Lambda\to\infty} (B-C)
  &=& \frac{g^2}{12\pi^2}
      \int_0^\infty {\rm d}l
   \left[
       \frac{8E_l^3}{m^4}
       -\frac{8l^3}{m^4}
       -\frac{12E_l}{m^2}
       +\frac{3}{E_l}
       +\frac{m^2}{E_l^3}
   \right]
\nonumber\\
  &=& \frac{g^2}{12\pi^2m^4}
      \int_0^\infty {\rm d}l
   \, \frac{{\rm d}}{{\rm d}l}
   \left[
       2lE_l^3 - 2l^4 - 3m^2lE_l + \frac{m^4l}{E_l}
   \right]
\nonumber\\
  &=& \frac{g^2}{48\pi^2}.
\end{eqnarray}

The nonzero value of $B-C$ shows that the one-loop fermion
self-energy introduces a Lorentz-violating term at small momenta
even when the cut-off is made very large: Lorentz violation is
suppressed not by a power of $m/\Lambda$, but only by a factor of
the coupling. A convenient interpretation of this result uses the
language of effective field theories.  Observe that the Lorentz
violation caused by the $B-C$ term is equivalent at the one-loop
level to adding a term
\begin{equation}
 -i\xi \bar \psi \gamma_\mu W^\mu W^\nu \partial_\nu\psi
\end{equation}
to the Lagrangian.  That is, we could obtain the same results in
the $\Lambda \gg m$ region by dropping the $B-C$ term in the
self-energy (\ref{eq:Sigma.exp1}) and then adding this extra term
to the Lagrangian.  In this extra term $\xi=g^2/48\pi^2$, is the
coefficient calculated above, and $W^\mu=(1,\vec{0})$ covariantly
specifies the rest frame of the cutoff. Summarizing,  the fermion
self-energy contribution has the effect of feeding back terms of
large unnatural size into the Lagrangian, thus producing a strong
fine tuning problem when higher corrections are taken into
account.
\paragraph{The boson self energy \cite{CPSUV}}
Similar results can be obtained from the boson self energy. Since
the internal lines are both fermionic in this case, a more
symmetrical way of incorporating the physical cutoff is required.
This is accomplished by extending the fermionic propagators to \be
\frac{i}{\gamma k-m+i\epsilon} \longrightarrow \frac{i \,
{f(|{\vec k}| / \Lambda)}}{\gamma k- m +{\Xi({\vec
k})}+i\epsilon}. \ee The function  $\Xi({\vec k})$ takes into
account the modified dispersion relations of the theory to zeroth
order and its contribution will not be taken into account in the
following.

Consider only the physical momentum cutoff function and calculate
the boson self-energy ${\Pi(E,{\vec p}, \Lambda)}$, defined in
terms of the full propagator {${\Delta(E,{\vec p})}$ \be
\Delta(E,{\vec p}) = \frac{{-i\,f(|{\vec k}| / \Lambda)}}{{p^2 -
\mu^2 -{\Pi(E,{\vec p}, \Lambda)} -i\epsilon }} \ee The one loop
calculation is {\begin{eqnarray} {{\Pi(E, {\vec p},\Lambda)} }&=&{
-\frac{ ig^2 }{ 16\pi^4 }
         \int d^4k\,\,
         {f(|\vec{{k}}|/\Lambda)
         f(|{\vec{k}+\vec{p}}|/\Lambda)}}\nonumber \\
         &&\times {\frac{ {\rm Tr}[ (\gamma\cdot k+m) \, (\gamma\cdot (k+p)+m)
] }
              { (k^2-m^2+i\epsilon) \, [ (k+p)^2-m^2+i\epsilon ]
              }.}
              \label{olphotse}
\end{eqnarray} }
The result can be presented as an expansion in even powers of the
 momenta
 \be \Pi(E, {\vec p},\Lambda)=
 A + p^2 B +
            {\eta_{RC}} p^\mu p^\nu W_\mu W_\nu
     + {\hat \Pi}(E, {\vec p},\Lambda ), \label{photse}
\ee where ${\hat \Pi}$ is convergent when the regulator is removed
so that it contributes only with Lorentz violating terms which are
suppressed by powers of  $p^2/\Lambda^2$. In the frame where the
cutoff $\Lambda$ is defined, $W_\mu=(1, {\vec 0})$, we obtain \be
\eta_{RC}=\frac{1}{2}\left(\left( \frac{\partial ^{2}\Pi
(p)}{\partial p^{0}\partial p^{0}}\right) _{p=0}+\left(
\frac{\partial ^{2}\Pi (p)}{\partial p^{1}\partial p^{1}}\right)
_{p=0} \right),\label{LIVPHOT} \ee from (\ref{photse}). The
calculation of the required pieces starts from (\ref{olphotse})
and after some algebra produces \be \left( \frac{\partial ^{2}\Pi
(p)}{\partial p^{0}\partial p^{0}}\right)
_{p=0}=-\frac{g^{2}}{32\pi ^{2}}\;\int_{0}^{\infty }dk\;k^{4}\;\frac{%
\;f^{2}}{E_{k}^{5}},\label{00}\ee \ba
\left(\frac{\partial ^{2}\,\Pi (p)}{\partial p^{i}\partial p^{j}}\right)_{p=0}&=&-\delta ^{ij}\frac{%
g^{2}}{48\pi ^{2}}\;\left[ \;\int_{0}^{\infty }k^{2}dk\;\left(
ff^{\prime \prime }\frac{k^{2}}{\Lambda ^{2}E_{k}^{3}}+ff^{\prime
}\frac{k}{\Lambda E_{k}^{5}}\left[ k^{2}+4m^{2}\right]\right.
\right. \nonumber \\
&& \left. \left. -f^{2}\frac{1}{2E_{k}^{7}}\left[
3k^{4}+10m^{2}k^{2}-3m^{4}\right] \right) \right]. \label{II}\ea
Here $f=f(k/\Lambda)$ and the derivatives, denoted by primes, are
with respect to the argument of $f$. As usual we have
$E_k=\sqrt{k^2+m^2}$. Each expression (\ref{00}) and (\ref{II}) is
logarithmically  divergent, but their combination (\ref{LIVPHOT})
is finite. Thus we can estimate the leading contribution to
$\eta_{RC}$ by setting $f=1$ inside the integrals with no
derivatives with respect to $f$. The terms containing derivatives
are handled  by assuming that the  contributions of $f',\, f''$
are sharply localized in the region $x=k/\Lambda\approx 1$. This
implies that we neglect the mass contributions $m^2<< k^2\approx
\Lambda^2$ in the corresponding integrals. Also, we set equal to
zero the resulting boundary terms which include $f'$ in the
required integrations by parts. The final result is \be \eta_{RC}
= \frac{g^2} {12\pi^2}
       \left[ 1 + 2 \int_0^{\infty} dx\,  x \, f'(x)^2 \right],
\ee which is estimated in the range $\eta_{RC} \geq 10^{-3} $
using standard model couplings. On the other hand, ${\eta_{RC}}$
can be interpreted as a correction {${\delta c }$}  to the boson
(photon) speed, for which extremely  tight bounds exists
$$ \frac{\eta_{RC}}{2} =\frac{\delta c}{c}= \leq 10^{-20}. $$

The two examples presented here show that radiative corrections in
preferred frames associated to space granularity, modelled by the
physical cutoff function (\ref{CUTOFFF}), induce  LIV
contributions which are enormously increased with respect to the
already established bounds. That is to say, they are not
suppressed by factors $1/M$ as expected initially. In this way, a
naturalness problem arises and a fine tuning problem appears when
considering higher order radiative corrections. One possibility to
guarantee the stability of dimension three and four LIV
contributions is via the introduction of a custodial symmetry.
This option has been explored using supersymmetry in
Ref.\cite{NIBBPOS}.

An alternative way of dealing with radiative corrections in LIV
processes can be found in Ref.\cite{ALFRC}.
\subsection{Radiation in Lorentz violating electrodynamics}
\label{ssec:radliv} The observation of $100\, MeV$ synchrotron
radiation from the Crab Nebula has recently been used to impose
stringent limits upon the parameters describing a modified
electrodynamics embodied in Maxwell equations, which together with
the corresponding coupled equations for the charges, get
correction terms which are linear in the Planck length
\cite{JLM1}. Such bounds were based on a set of very reasonable
assumptions on how some of the standard results of synchrotron
radiation extend to the Lorentz non-invariant situation. This
caused some controversy in the literature \cite{CONTROV}.
Moreover, an assessment of such assumptions requires the
introduction of specific dynamical models. One of them is the
Myers-Pospelov (MP) effective theory at the classical level, which
parameterizes LIV using dimension five operators Ref. \cite{MP}. A
detailed description of synchrotron radiation in this model has
been presented in Refs.\cite{UM}, which we review in this
subsection and that has motivated the general point of view
summarized in subsection \ref{ss102}, given in more detail in
Ref.\cite{UM1}.

The study of radiation in LIV electrodynamics  constitutes an
interesting problem on its own whose resolution will subsequently
allow the use of additional observational information to put
bounds upon the correction parameters. For example we have in mind
 polarization measurements from cosmological sources. The case of
gamma ray bursts  has recently become increasingly relevant
\cite{CB}, although it is still at a controversial stage
\cite{CONTR}.

Our calculation of synchrotron radiation in modified
electrodynamics rest heavily on previous work  reported in
Refs.\cite{SCH49,SCHWBOOK,SCHWANNP,ERBER,EWL}.

A partial list of previous studies in electrodynamics
incorporating LIV via dimension three and four operators is given
in Ref.\cite{ALL}
\subsubsection{Synchrotron radiation in the Myers-Pospelov  model}

This model parameterizes LIV using dimension five operators both
in the matter and electromagnetic sectors. There is also a
preferred frame four velocity $V^\mu$, which is not a dynamical
field. As usual the model exhibits passive (observer) Lorentz
covariance, which means that the fields and the four-velocity
$V^\mu$ transform as tensors when going from one observer frame to
another. On the other hand, in each frame we violate active
(particle) Lorentz transformations; that is to say  we have the
non-dynamical physical field $ V^\mu$ in the action,  in analogy
to the physics going on in the presence of an external magnetic
field which violates active rotational invariance, for example.

\paragraph{The charge sector}

The dynamics of a classical charged particle of mass $\mu$ can be
obtained from the action for a scalar charged field. In this case
the Myers-Pospelov action is
\begin{equation}
S_{MP}=\int d^{4}x\;\left[ \partial _{\mu }\varphi ^{\ast
}\partial ^{\mu }\varphi -\mu ^{2}\varphi ^{\ast }\varphi
+i{\tilde \eta}\, \varphi ^{\ast }\left( V\cdot \partial \right)
^{3}\varphi \right] ,
\end{equation}
with the notation $V\cdot \partial =V^{\mu }\,\partial _{\mu }$.
In momentum space, where we write $\varphi (x)=\varphi _{0}\,\exp
i(p^{0}t-\mathbf{p\cdot x})$ and in the reference frame where
$V^{\alpha }=\left( 1,\mathbf{0 }\right) $, the modified
dispersion relation becomes
\begin{equation}
\left( p^{0}\right) ^{2}+\tilde{\eta}\left( p^{0}\right)
^{3}=\mathbf{p} ^{2}+\mu ^{2}  \label{EXDR}.
\end{equation}
To make contact  with the  results in Ref.\cite{JLM1} it is
necessary  to rewrite $ \tilde{\eta}=-{\eta}/{M},\, \eta <0, $
where $\eta $ is a dimensionless constant and  $M$ is  a mass
scale characterizing the Lorentz symmetry breaking, which is
usually, but not necessarily, identified with the Plank mass. The
equation (\ref{EXDR}) is an exact relation in $\tilde{\eta}$. From
here we obtain the Hamiltonian for a massive particle to second
order in $ \tilde{\eta}$
\begin{equation}
p^0=H=\left( \mathbf{p}^{2}+\mu ^{2}\right)
^{1/2}-\frac{1}{2}\tilde{\eta}\left( \mathbf{p}^{2}+\mu
^{2}\right) +\frac{5}{8}\tilde{\eta}^{2}\left( \mathbf{p} ^{2}+\mu
^{2}\right) ^{3/2}+O(\tilde{\eta}^{3}).
\end{equation}
In the following we consider the interaction of a particle having
charge $q$ with a static constant magnetic field $\mathbf{B}={\vec
\nabla}\times{\vec A}$. The standard minimal coupling produces the
Hamiltonian
 \ba H&=&\left[ \left(
\mathbf{p-}\frac{q}{c}\mathbf{A}\right) ^{2}+\mu ^{2}\right]
^{1/2}-\frac{1}{2}\tilde{\eta}\left[ \left(
\mathbf{p-}\frac{q}{c}\mathbf{A} \right) ^{2}+\mu ^{2}\right]
\nonumber \\
&&+\frac{5}{8}\tilde{\eta}^{2}\left[ \left(
\mathbf{p-}\frac{q}{c}\mathbf{A}\right) ^{2}+\mu ^{2}\right]
^{3/2}, \label{HAMCERO}\ea to order $O(\tilde{ \eta}^{3})$.   Here
$c=3\times 10^{10}\, cm\, s^{-1}$ denotes the uncorrected velocity
of light in vacuum. In the following  we set $c=1$. Observe that
the dispersion relation (\ref{EXDR}) provides the exact inversion
\begin{equation}
\left( \mathbf{p}-q\mathbf{A}\right) ^{2}=\left( 1+\tilde{\eta}
E\right) E^{2}-\mu ^{2},  \label{PMA}
\end{equation}
with $E$ being the energy of the particle. The Hamilton equations
arising from (\ref{HAMCERO}) yield the acceleration
\begin{equation}
\mathbf{\ddot{r}}=\frac{q}{E}\left(1-\frac{3}{2}\tilde{\eta}E+\frac{9}{4}
\tilde{\eta}^{2}E^{2}\allowbreak \right) \left( \mathbf{v\times
B}\right) .
\end{equation}
As in the usual case, this means that the magnitude $|\mathbf{v|}$
of the particle velocity is constant and that the projection of
the trajectory  in a plane perpendicular to $ \mathbf{B}$ is a
circular orbit with a Larmor frequency
\begin{equation}
\omega
_{0}=\frac{|q|B}{E}\left(1-\frac{3}{2}\tilde{\eta}E+\frac{9}{4}
\tilde{\eta}^{2}E^{2}\allowbreak \right) .  \label{OMEGA0}
\end{equation}
In general the motion is an helix with pitch angle (the angle
between the velocity and the magnetic field) $\alpha $. We
restrict ourselves to the motion in the plane perpendicular to the
magnetic field, i.e. $\alpha=\pi/2$. The solution to the equations
of motion can be written as
\begin{equation}
\mathbf{r(}t)\mathbf{=}\left( \frac{\beta}{\omega _{0}}\cos \omega
_{0}t,\ \frac{\beta}{\omega _{0}}\sin \omega _{0}t,\ 0 \right),
\label{rt}
\end{equation}
where we emphasize that we are using the standard definition\
$\beta =| \mathbf{v|}/c$. The velocity is
\begin{equation}
\mathbf{v(}t)\mathbf{=}(-\beta \sin \omega _{0}t,\;\beta \cos
\omega _{0}t,\;0).  \label{vt}
\end{equation}
The  equation (\ref{rt}) identifies $ R={\beta }/{\omega _{0}}$ as
the Larmor radius of the orbit. The Lorentz factor $\gamma$ is
given by
\begin{equation}
\frac{1}{\gamma^2}=1-\beta ^{2}=\frac{\mu ^{2}}{E^{2}}\left[
1+2\frac{\tilde{\eta}E^{3}}{\mu
^{2}}-\frac{15}{4}\frac{\tilde{\eta}^{2}E^{4}}{\mu
^{2}}+O(\tilde{\eta}^{3}) \right] ,  \label{UMB2}
\end{equation}
where the range of energies to be considered is such that $
\frac{\mu }{E}<<1, \, \tilde{\eta}E<<1$.

According to the preceding analysis, the current for a charged
particle moving in the circular motion is
\begin{equation}
\mathbf{j}(t,\mathbf{r})=q\delta ^{3}(\mathbf{r}-\mathbf{r}(t))\,\mathbf{v}%
(t),  \label{CURRENT}
\end{equation}%
where $\mathbf{r}(t)$ and $\mathbf{v}(t)$ are given in Eqs.
(\ref{rt}) and (\ref{vt}) respectively.
\paragraph{The electromagnetic sector}
The corresponding action is \cite{MP} \be S_{photon}=\int
d^{4}x\left[ -\frac{1}{4}F_{\mu \nu }F^{\mu \nu }-4\pi \, J^{\mu
}A_{\mu } + {\tilde \xi} \left( {V^{\alpha }}F_{\alpha \delta
}\right) {(V^\nu}
\partial_\nu )({V_{\beta }}\tilde{F}^{\beta \delta })\right].
\ee
We choose to work in the rest frame $V^\mu=(1,\mathbf{0})$
where the modified Maxwell equations are ($c=1$) \ba &&\nabla
\cdot \mathbf{B}=0,  \qquad\quad  \mathbf{\nabla \times
E+}\frac{\partial \mathbf{B}}{\partial t}=0,  \nonumber
\\
&&\nabla \cdot \mathbf{E}=4\pi\rho, \qquad  -\frac{\partial
\mathbf{E}}{\partial t}+\nabla \times \mathbf{B}+{\tilde \xi}
\partial _{0}\left( -\nabla \times \mathbf{E}+\partial _{0}\mathbf{B}%
)\right)=4\pi \mathbf{j}. \nonumber \\ \label{MP}
\ea%
For future purposes it is convenient to define $\tilde \xi=\xi/M$,
where $\xi$ is a dimensionless parameter.  Introducing the
standard potential fields $A^{\mu }=\left( \phi ,\mathbf{A}
\right) $ in the Coulomb gauge we have \ba
\phi &=&-4\pi \frac{1}{\nabla ^{2}}\rho , \\
\frac{\partial ^{2}\mathbf{A}}{\partial t^{2}}-\nabla ^{2}\mathbf{A}+2\tilde{%
\xi}\nabla \times \frac{\partial ^{2}\mathbf{A}}{\partial t^{2}}&
=&4\pi
\,\left( \mathbf{j-}\nabla \frac{1}{\nabla ^{2}}\nabla \cdot \mathbf{j}%
\right) \equiv 4\pi \,\mathbf{j}_{T},  \label{ema}
\ea%
where the electric and magnetic fields  reduce to
\begin{equation}
\mathbf{E}=-\frac{\partial \mathbf{A}}{\partial
t},\;\;\;\;\mathbf{B}=\nabla \times \mathbf{A},
\end{equation}
in the radiation approximation.

The energy momentum tensor $T_{\mu \nu }$ for this modified
electrodynamics
is given by%
\ba
T_{\;0}^{0}& =\frac{1}{8\pi }(\mathbf{E}^{2}+\mathbf{B}^{2})-\frac{\tilde{\xi%
}}{4\pi }\mathbf{E\,\cdot \,}\frac{\partial \mathbf{B}}{\partial
t},
\label{ed} \\
\mathbf{S}& =\frac{1}{4\pi }\mathbf{E}\times \mathbf{B}-\frac{\tilde{\xi}}{%
4\pi }\mathbf{E}\times \frac{\partial \mathbf{E}}{\partial t},
\label{pv}
\ea%
which are exact expressions in $\tilde{\xi}$ and satisfy  the
usual conservation equation outside the sources.

To solve the equation of motion for $\mathbf{A}$, Eq. (\ref{ema}),
we can go to the momentum space with the convention
\begin{equation}
F(t,\mathbf{r})=\int \frac{d^{4}k}{\left( 2\pi \right) ^{4}}e^{-i\omega t+i%
\mathbf{k}\cdot \mathbf{r}}F(\omega ,\mathbf{k}).
\end{equation}%
The different types of Fourier transforms are specified by the
corresponding arguments. For example, if $F(t,\mathbf{r)}$ denotes
the function in space-time, $F(\omega ,\mathbf{r)}$ denotes the
Fourier transformed function to frequency space, while $F(\omega
,\mathbf{k})$ denotes the Fourier transformed function to
frequency and momentum spaces. In this way Eq. (\ref{ema}) reduces
to
\begin{equation}
\left( -\omega ^{2}+k^{2}-2i\tilde{\xi}\omega
^{2}\,\mathbf{k}\,\times \right) \mathbf{A(}\omega
\mathbf{,k)}=4\pi \,\mathbf{j}_{T}\mathbf{(}\omega ,\mathbf{k)}.
\end{equation}%
This equation can be diagonalized using the circular polarization
basis, with $\lambda =\pm 1$, giving
\begin{equation}
\left( -\omega ^{2}+k^{2}\mp 2\tilde{\xi}\omega ^{2}k\right)
\mathbf{A}^{\pm }\mathbf{(}\omega \mathbf{,k)}=4\pi
\mathbf{j}_{T}^{\pm }\mathbf{(}\omega \mathbf{,k)}.
\label{DEFPOL}
\end{equation}
The components  $C^\lambda_i$ of any vector ${\vec C}$ in the
polarization basis associated to the direction ${\vec k}$ are
 \be C_{i}^{\lambda }=P_{ik}^{\lambda }C_{k}\,,\qquad
P_{ik}^{\lambda }=\frac{1}{2}\left(
\delta _{ik}-{\hat{k}}_{i}{\hat{k}}_{k}+ \lambda i\epsilon _{ijk}{\hat{k}}%
_{j}\right) .  \label{PROY} \ee The simplest way to proceed is by
introducing the total retarded Green function \be [G_{ret}(\omega
,\mathbf{k})]_{ik}=\sum_{\lambda}\, P_{ik}^{\lambda }\, \left.
\frac{1}{ k^{2}-\lambda 2\tilde{\xi}\omega ^{2}k-\omega ^{2}
}\right\vert _{\omega \rightarrow \omega +i\epsilon }.
\label{GRETGF0} \ee and to calculate \be [G_{ret}(\omega , {\vec
r}- {\vec r}^{\prime })]_{ik}= \int
\frac{d^{3}\mathbf{k}}{\left( 2\pi \right) ^{3}}e^{i\mathbf{k}\cdot (\mathbf{%
r-r}^{\prime })}[G_{ret}(\omega ,\mathbf{k})]_{ik}.\ee After some
rearrangements the final integration over $dk$ can be performed by
the method of residues in the complex plane. Finally one can
identify back the polarization components of the total Green
function as
\begin{equation}
G_{ret}^{\lambda }\mathbf{(}\omega \mathbf{,r-r}^{\prime
}\mathbf{)=}\frac{1}{%
4\pi R}\frac{n_\lambda(z)}{\sqrt{1+z^{2}}}\,e^{in_\lambda(z)\omega
R}, \label{GRETGF}
\end{equation}%
where $R=|\mathbf{r}-\mathbf{r}^{\prime }|$.  Here we have
introduced the polarization-dependent refraction index
$n_\lambda(z)$
\begin{equation}
n_\lambda(z)=\sqrt{1+z^{2}}+\lambda z,\;\;\mathbf{\
}z=\tilde{\xi}\omega . \label{REFIND}
\end{equation}%
In this way, the fields $\mathbf{A^{\lambda }}$ in Eq.
(\ref{DEFPOL}) have well defined phase velocities $v_{\lambda
}=1/n_\lambda(z)$ and this situation can be described as the
propagation of photons in a dispersive birefringent medium.

The Green functions (\ref{GRETGF}) determine the corresponding
potentials with the standard replacements $1/R\simeq 1/|
\mathbf{r}|\equiv 1/r$ in the denominator, together with the
following expansion of the phase $n(\lambda z)\omega R$
\begin{equation}
n_\lambda(z)\omega\left\vert
\mathbf{r}-\mathbf{r}^{\prime}\right\vert
\simeq\omega r\left[ 1-\frac{\mathbf{\hat n}\cdot\mathbf{r}^{\prime}}{r}+\lambda%
\tilde{\xi}\omega-\lambda\tilde{\xi}\omega\frac{\mathbf{\hat n}\cdot\mathbf{r}%
^{\prime}}{r}+\frac{1}{2}\left( \frac{r^{\prime}}{r}\right)
^{2}\right] , \label{PHASE}
\end{equation}
where $\mathbf{\hat{n}}=\mathbf{r} /r $ is the direction of
observation. Notice that we are interested in the radiation
approximation of the phase (\ref{PHASE}), which means that the
subdominant terms of order $(r^{\prime}/r)^2$ or higher are
neglected. Consistency demands that the terms proportional to the
LIV parameter $\tilde{\xi}$ are larger than the neglected one in
order to properly include them in this phase. Our general results
are presented in this full far-field approximation.

Using (\ref{GRETGF}) we  finally get
\begin{equation}
\mathbf{A}^{\lambda }(\omega ,\mathbf{r})=\frac{1}{r}\frac{n_\lambda(z)}{%
\sqrt{1+{z^{2}}}}e^{in_\lambda(z)\omega r}\mathbf{j}^{\lambda
}(\omega, \mathbf{{k}_{\lambda }}) \label{APOL}
\end{equation}%
in the radiation approximation. The fields $\mathbf{A}^{+}(\omega
, \mathbf{r})$ and $\mathbf{A}^{-}(\omega ,\mathbf{r})$ correspond
to right and left circular polarization respectively. Let us
emphasize that the momenta
\begin{equation}
\mathbf{k}_{\lambda }=n_\lambda(z)\ \omega \
\mathbf{\mathbf{\hat{n} }} \label{KLAMBDA}
\end{equation}
in Eq. (\ref{APOL}) are fixed in terms of the frequency and the
direction of observation. The full vector potential is given by
the superposition ${\vec A}(\omega ,\mathbf{r})={\vec A}^{+
}(\omega ,\mathbf{r})+{\vec A}^{-}(\omega ,\mathbf{r})$. Hence the
electric and magnetic fields are \ba \mathbf{B(}\omega
,\mathbf{r})&=& \frac{1}{r}\frac{\omega
}{\sqrt{1+z^{2}}}\,\sum_{\lambda=\pm}\,
\lambda\, n_\lambda^{2}(z)\, e^{in_\lambda(z)\omega r}%
\, \mathbf{j}^{\lambda}\mathbf{(}\omega ,\mathbf{k}_\lambda),\nonumber \\
\mathbf{E(}\omega ,\mathbf{r})&=& \frac{1}{r}\frac{i\omega
}{\sqrt{1+z^{2}}} \,\sum_{\lambda=\pm}\, n_\lambda(z)\,
e^{in_\lambda(z)\omega r}\, \mathbf{j}^{\lambda}\mathbf{(}\omega
,\mathbf{k}_\lambda). \label{EFIELD}
\ea%
Note that, contrary to the standard case where
$\mathbf{\hat{n}\times E} \propto \mathbf{B}$, here we have
\begin{equation}
\mathbf{\hat{n}\times E}(\omega
,\mathbf{r})=\frac{1}{\sqrt{1+z^{2}}}\left[ \mathbf{B}(\omega
,\mathbf{r})+iz\mathbf{E}(\omega ,\mathbf{r})\right] .
\end{equation}

The angular distribution of the power spectrum is defined as \be
\frac{d^{2}P(T)}{d\omega d\Omega}, \ee
 where $P(T)$ is the radiated
power at time $T$ into the solid angle $d\Omega$. We can compute
the total energy emitted in
terms of the Poynting vector (\ref{pv})%
\begin{equation}
E=\int_{-\infty }^{\infty }dt\;\mathbf{n\cdot
S(}t,\mathbf{r)\;}r^{2}d\Omega .
\end{equation}%
This last expression can be rewritten introducing the Fourier
transform of the Poynting vector, \ba
E=\int_{0}^{\infty }d\omega \int d\Omega \;\frac{d^{2}E}{d\Omega d\omega }%
=\int_{0}^{\infty }\frac{d\omega }{2\pi }\left[ \mathbf{n\cdot S(}\omega ,%
\mathbf{r)+n\cdot S(-}\omega ,\mathbf{r)}\right]
r^{2}d\Omega,\nonumber \\
\ea%
and allows us to obtain the angular distribution of the energy
spectrum
\begin{equation}
\frac{d^{2}E}{d\Omega d\omega }=\frac{r^{2}}{2\pi }\left[ \mathbf{n\cdot S(}%
\omega ,\mathbf{r)+n\cdot S(-}\omega ,\mathbf{r)}\right] ,
\label{eo}
\end{equation}%
from where the angular distribution of the power spectrum can be
identified as
\begin{equation}
\frac{d^{2}E}{d\Omega d\omega }=\int_{-\infty}^{+\infty}
dT\;\frac{d^{2}P(T)}{dwd\Omega }. \label{ep}
\end{equation}%
Writing the Poynting vector (\ref{pv}) in terms of the polarized
potentials we obtain the intermediate result
\begin{equation}
\frac{d^{2}E}{d\Omega d\omega }=\frac{r^{2}\omega ^{2}}{4\pi ^{2}}\sqrt{%
1+z^{2}}\left[ \mathbf{A}_{-}(-\omega ,\mathbf{r})\cdot \mathbf{A}%
_{+}(\omega ,\mathbf{r})+\mathbf{A}_{-}(\omega ,\mathbf{r})\cdot \mathbf{A}%
_{+}(-\omega ,\mathbf{r})\right] .
\end{equation}%
Next we  express the products $\mathbf{A}_{\mp }(-\omega
,\mathbf{r} )\cdot \mathbf{A}_{\pm }(\omega ,\mathbf{r})$ in terms
of the current $ \mathbf{j}(\omega ,\;\mathbf{k})$ via the
relation (\ref{APOL}). Using the properties of the fields in the
polarization basis,  together with the general relation $
j_{k}(-\omega ,-\mathbf{k})=j_{k}^{\ast }(\omega ,\mathbf{k})$, we
obtain
\begin{eqnarray}
\frac{d^{2}E}{d\Omega d\omega } &=&\frac{1}{4\pi ^{2}}\frac{\omega ^{2}}{%
1+z^{2}} \left[\frac{}{}  n_+^{2}(z)\ j_{k}^{\ast }\left( \omega
,{\mathbf k}_+\right) \;P_{kr}^{+}\;j_{r}(\omega ,{\mathbf k}_+%
) \right. \nonumber \\
&& \left. +n_-^{2}(z)\;j_{k}^{\ast }(\omega, {\mathbf k}_-)%
\;P_{kr}^{-}\;j_{r}\left( \omega,{\mathbf k}_-\right)\frac{}{} %
\right] .  \label{es}
\end{eqnarray}
In order to identify the angular distribution of the power
spectrum we need to introduce the time dependence via the
corresponding inverse Fourier transform. Each contribution in Eq.
(\ref{es}) is of the type
\begin{equation}
C\left( \omega \right) =j_{k}^{\ast }\left( \omega
,\mathbf{k}\right) \ X_{kr}\;j_{r}\left( \omega ,\mathbf{k}\right)
=\int_{-\infty}^{+\infty} dt\ dt^{\prime
}e^{-i\omega \left( t-t^{\prime }\right) }j_{k}^{\ast }\left( t,\mathbf{k}%
\right) \ X_{kr}\;j_{r}\left( t^{\prime },\mathbf{k}\right) .
\end{equation}%
Changing to new time variables $\tau =t-t^{\prime }$ and $T=\left(
t+t^{\prime }\right) /2$ we get%
\begin{equation}
C\left( \omega \right) =\int_{-\infty}^{+\infty} dT\int_{-\infty
}^{\infty }d\tau e^{-i\omega \tau }j_{k}^{\ast }\left( T+\tau
/2,\mathbf{k}\right) \ X_{kr}\;j_{r}\left( T-\tau
/2,\mathbf{k}\right) .
\end{equation}%
Inserting this last relation in Eq. (\ref{es}) and comparing with
Eq. (\ref {ep}) we obtain the final expression for the angular
distribution of the radiated power spectrum
\begin{eqnarray}
\frac{d^{2}P(T)}{d\omega d\Omega}&=&\frac{1}{4\pi ^{2}}\frac{\omega ^{2}}{%
\sqrt{1+z^{2}}}\int_{-\infty }^{\infty }d\tau \ e^{-i\omega \tau
}\times
\nonumber \\
&&\sum_{\lambda} \frac{}{} n_\lambda^{2}(z)j_{k}^{\ast } \left(
T+\tau /2,{\mathbf k}_\lambda \right)
P_{kr}^{\lambda}\;j_{r}\left( T-\tau /2,{\mathbf k}_\lambda
\right), \label{poo}
\end{eqnarray}
as the sum of the contributions of both circular polarizations.
\paragraph{Synchrotron radiation}
This corresponds to the choice (\ref{vt}) for the velocity of the
current (\ref{CURRENT}) in the general expression (\ref{poo}).
Following the method of Ref.\cite{SCHWBOOK} the time-averaged
angular distribution of the radiated power spectrum  is \ba
&&\left\langle \frac{d^{2}P(T)}{d\omega d\Omega}\right\rangle
=\sum
_{\lambda=+,-}\,\sum_{m=0}^{\infty}\delta({\omega}-\omega_{m})\frac
{dP_{m,\,\lambda}}{d\Omega}, \quad \omega_{m}=m\omega_{0},
\quad z_{m}=\tilde{\xi}\omega_{m}, \label{HARMEXP} \nonumber \\
&&\frac{dP_{m,\,\lambda}}{d\Omega}=\frac{\omega_{m}^{2}q^{2}}{4\pi}\frac
{1}{\sqrt{1+z_{m}^{2}}}\left[  \lambda \beta n_\lambda(
z_{m})J_{m}^{\,\prime }(W_{\lambda
m})+\,\cot\theta\,J_{m}(W_{\lambda m})\right]^{2},\nonumber \\
\label{ADMH}%
\ea written as a sum over the contribution of the harmonics
$\omega_m$. Here, the average $\left\langle \right\rangle $ is
taken with respect to the macroscopic time $T$ and $J_{m}$,
$J_{m}^{\,\prime}$ denote the Bessel functions and their
derivatives respectively. The argument of the Bessel functions is
$ W_{\lambda m}=m\,n_\lambda( z_{m})\beta\sin\theta$.

We also have calculated the total averaged and integrated
power radiated into the $m$-th harmonic%
\ba P_{m}&=&\frac{q^{2}\beta^{2}\omega_{m}}{R
\,\sqrt{1+z_{m}^{2}}}\,\sum_{\lambda=\pm
}n_\lambda( z_{m})\left[J_{2m}^{\,\prime}(2m\;n_\lambda( z_{m}%
)\beta)\right. \nonumber \\
&& \left. -\frac{1-\beta^{2}n_\lambda^{2}(
z_{m})}{2\beta^{2}n_\lambda^{2}( z_{m})}  \int
_{0}^{2mn_\lambda( z_{m})\beta}dx\;J_{2m}(x)\right]  , \label{PMDO}%
\ea which clearly indicates the contribution of each polarization
$P_{\lambda m}$. The above result is exact in $z_{m}$ and the
parity-violating contribution has vanished after the angular
integration.
\begin{table}
\centering \caption{Data of some relevant astrophysical objects}
\label{tab:3}
\begin{tabular}{llllllll}
\hline\noalign{\smallskip} ${\rm Object}\quad \quad $ & $\ \  r \,
(l.y.) \quad$ &$\gamma\quad$\quad&$\ \ { B \
(Gauss)}\quad$&$\omega_{c} (GeV)\quad$&$\omega_0 (GeV)\quad$
&$\  m\quad$& $\ \  m/\gamma $\\
\noalign{\smallskip}\hline\noalign{\smallskip}
CRAB &$\ \ \  10^4$&$10^9$&$\qquad \ 10^{-3}$& $\ \ \ 10^{-1}$&$\ \  10^{-30}$&$  10^{29}$& $ \ \   10^{20}$\\
  $(Mkn\, 501)_p$&$\ \ \  10^8$&$10^{11}$&$\qquad \ 10^{2}$&$\ \ \ 10^4$&$\ \  10^{-29}$&$  10^{33}$&$\ \    10^{22}$\\
  $(Mkn\, 501)_e$&$\ \ \  10^8$&$10^{11}$&$\qquad \ 10^{-1}$&$\ \ \ 10^4$&$\ \  10^{-29}$&$  10^{33}$&$\ \    10^{22}$\\
$GRB \, 021006$&$\ \ \   10^{10}$&$10^5$&$\qquad \ 10^{4}$&$\ \ \ 10^{-3}$&$\ \  10^{-18}$&$  10^{15}$&$\ \    10^{10}$\\
\noalign{\smallskip}\hline
\end{tabular}
\end{table}

In Table \ref{tab:3} we present  a rough estimation of the
relevant parameters associated  with some observed cosmological
objects. There $r\,[l.y.]$ is the distance of the object to the
earth, $\gamma$ is the Lorentz factor of the charged particles at
the cutoff frequency, $B\,[Gauss]$ is the average magnetic field
producing the synchrotron radiation, $\omega_c\,[GeV]$ is the
cut-off frequency and  $\omega _{0}\,[GeV]$ is the Larmor
frequency. In all cases the cut-off frequency $\omega_c$ has been
estimated from the radiation spectrum fitted by a synchrotron
model in the corresponding reference. This, together with the
magnetic field $B$ allows us to estimate the Lorentz factor
\begin{equation} \gamma=2.36\times 10^8
\sqrt{\frac{\omega_c[GeV]}{B[Gauss]}\frac{M}{m_e}},
\end{equation}
where $M$ is the mass of the charged particle. From the above we
further obtain the Larmor frequency
\begin{equation} \omega_0[GeV]=0.6 \times 10^{-17}\left(\frac{2m_e}{\gamma M}
\right) B[Gauss].
\end{equation}
In the case of CRAB Nebula we adopt the typical values given in
Ref. \cite{ATOYAN}. For Mkn 501 we consider two possible models
for synchrotron radiation where the emitter particles are either
protons \cite{AHARONIAN} or electrons \cite{KONOPELKO}. In the
latter case we  use the radius of the orbit $r'=1.5 \times 10^{15}
\, cm =1/\omega_0$ and the magnetic field to obtain the Lorentz
factor. Finally we consider the GRB021206 ($z \approx 1.25 $).
According to Ref. \cite{NAKAR} this object has a very compact core
with a radius of the order of $1 \, km$ and a magnetic field
$\approx 10^{12} \; Gauss$. The synchrotron emission region is
about $10^8 \, km$ from the core \cite{WAXMAN}, so that we
estimate the magnetic field to be $10^4 \; Gauss$ using the
transport law $B\,r = const.$. From Ref. \cite{HAJDAS} we take the
cut-off frequency to be $\omega_c=1 \; MeV$.

As indicated in Table \ref{tab:3}., the radiation of interest is
dominated by very high harmonics $10^{15}\leq m \leq 10^{30}$
exhibiting also large ratios of $m/\gamma$, typically in the range
$10^{10}\leq m/\gamma \leq 10^{22}$. The corresponding values for
$\gamma$ imply also $\beta \approx 1$. In this way, the high
harmonics present in the synchrotron radiation of these
astrophysical sources together with  the values of the $\gamma $
factors of the radiating charges highlight the relevance of the
large $m$ and large  $\gamma $ limit, with the constraint $\left(
m/\gamma \right) ^{2}\gg 1$, to study the induced Lorentz
violating effects.

In a similar way to the standard case \cite{SCHWANNP}, we obtain
\ba
 P_{\lambda m}&=& \frac{q^{2}m\omega_{0}}{\sqrt{3}\pi
R}\frac{1}{1+n^{2}(\lambda {z_m})}\left\{  \int_{m/{\tilde{m}_{\lambda c}}%
}^{\infty}dx\, \left(\frac{3}{2{{\tilde
m}_{\lambda c}}}\right)^{2/3}\,K_{5/3}\left(x\right)\right. \nonumber \\
&&
 \left.-2 \left(\frac{3}{2{{\tilde m}_{\lambda c}}}\right)^{4/3}\,
 K_{2/3}\left(
\frac{m}{{\tilde{m}_{c}}}\right)  \right \},\nonumber \ea for the
integrated power in the $m^{\rm th}$ harmonic. Here $K_{p/q}$
denote  the Macdonald functions (Bessel functions of fractional
order). A first consequence of this approximation is the
appearance  of the cutoff frequency $\omega_{\lambda
c}=\tilde{m}_{\lambda c}\,\omega_0$ with
\begin{equation}
\tilde{m}_{\lambda c}=\frac{3}{2}\left[  1-\beta^{2}(E)\,n_\lambda^{2}({z_{m}%
})\right]  ^{-3/2}. \label{HCOFF}%
\end{equation}
This name arises because  for $m > \tilde{m}_{\lambda c}$ the
total power decreases as \be P_{\lambda m} \approx
\exp(-m/{\tilde{m}_{\lambda c}}). \ee Within the same large-$m$
approximation, the integrated power in the $m$-th harmonic can be
expanded to second order in
${\tilde{\xi}}$ yielding%
\ba
P_{m}&=&\frac{q^{2}\omega}{\sqrt{3}\pi R \gamma^{2}}\left[  \frac{m_{c}}%
{m}{\kappa}\left(  \frac{m}{m_{c}}\right)  -\frac{2}{\gamma^{2}}%
K_{2/3}\left(  \frac{m}{m_{c}}\right) \right.\nonumber\\
&&\left. +2\left(  \frac{{\tilde{\xi}}%
\,m\omega\beta}{\gamma}\right)  ^{2}K_{2/3}\left(
\frac{m}{m_{c}}\right)
\right]  , \label{PMEXP}%
\ea where $m_{c}=3\gamma^{3}/2$ and
${\kappa}(x)=x\int_x^{\infty}dy\,K_{5/3}(y)$ is the so called
brems\-strah\-lung function.

Let us notice  the appearance of the combination ${\tilde{\xi
}}\,\omega m/\gamma=\xi(\omega/M_{P})(m/\gamma)$ as the expansion
parameter in (\ref{PMEXP}). Here we take $M=M_P$. As can be seen
from Table \ref{tab:3} this is not necessarily a small number,
which signals the possibility that such corrections might be
relevant in setting bounds upon $\tilde{\xi}$. This rather
unexpected effect is due to the amplifying factor $ m/\gamma$.
Similar results have been obtained in calculations of the
synchrotron radiation spectra in the context of non-commutative
electrodynamics \cite{Castorina}.

Another possibility for observable effects due to ${\tilde{\xi}}$
is to look
at the averaged degree of circular polarization%
\begin{equation}
\Pi_{\odot}=\frac{\left\langle
P_{+}(\omega)-P_{-}(\omega)\right\rangle
}{\left\langle P_{+}(\omega)+P_{-}(\omega)\right\rangle }, \label{CIRCPOL}%
\end{equation}
where $P_{\lambda}(\omega)$ is the total power distribution per
unit frequency
and polarization $\lambda$, so that $P_{\lambda}(\omega)=P_{m\lambda}%
/\omega_{0}$. The average here is calculated with respect to an
energy distribution of the relativistic electrons, which we take
to be $N(E)dE=CE^{-p}dE$, in some energy range $E_{1}<E<E_{2}$,
chosen  as $E_1=0$ and $E_2 \rightarrow \infty$
for simplicity. The result is%
\be \Pi_{\odot}={\tilde{\xi}}\omega\left(
\frac{\mu\omega}{qB}\right) \,\Pi(p),
\label{CPEXP}%
\ee
\begin{equation}
\Pi(p)=\frac{(p-3)\left(  3p-1\right)  }{3\left(  3p-7\right)
}\,\frac {(p+1)}{(p-1)}\frac{\Gamma\left(
\frac{p}{4}+\frac{13}{12}\right)
\Gamma\left(  \frac{p}{4}+\frac{5}{12}\right)  }{\Gamma\left(  \frac{p}%
{4}+\frac{19}{12}\right)  \Gamma\left(
\frac{p}{4}+\frac{11}{12}\right)
},\quad p>7/3. \label{PIP}%
\end{equation}
Again, we have the presence of an amplifying factor in Eq.
(\ref{CPEXP}), given by $(\mu\omega/qB)$, which is independent of
the form of $\Pi(p)$ and not necessarily a small number.  An
estimation of this factor in the zeroth-order approximation
(${\tilde \xi}=0={\tilde \eta}$), which is appropriate in
 (\ref{CPEXP}), yields $(\mu\omega/q B )=\omega/(\omega_0
\gamma)=m/\gamma$. The expression (\ref{CPEXP}) is analogous to
the well-known average of the degree of linear polarization
$\Pi_{LIN}=(p+1)/(p+7/3)$, under the same energy distribution for
the electrons \cite{RL}.

Finally in Table \ref{tab:4}. we have  estimated the contributions
to the different pieces in the phase (\ref{PHASE}) of the Green
function, for each astrophysical object. Here we have set $\xi=1$
and $M=M_P$.
\begin{table}
\centering \caption{The far field approximation} \label{tab:4}
\begin{tabular}{lllll}
\hline\noalign{\smallskip} ${\rm Object}\quad \quad $ & $\qquad
{r'}/{r}
 \quad$ &$\quad \omega_c/M_P\quad$\quad&$ \quad ( r'/r)(\omega_c/M_P)\quad$&$\quad \ \ (r'/r)^2$\\
\noalign{\smallskip}\hline\noalign{\smallskip}
CRAB &$\qquad  10^{-6}$&$\qquad 10^{-20}$&$\qquad \quad 10^{-26}$& $\qquad 10^{-12}$\\
  $Mkn\, 501$&$\qquad  10^{-11}$&$\qquad 10^{-15}$&$\qquad \quad 10^{-26}$&$\qquad  10^{-22}$\\
$GRB \, 021006$&$\qquad   10^{-24}$&$\qquad 10^{-22}$&$\qquad \quad 10^{-46}$&$\qquad  10^{-48}$\\
\noalign{\smallskip}\hline
\end{tabular}
\end{table}
The extreme case is \be {\left\vert \tilde \xi \omega \right\vert
\frac{r'}{r}} < {\left\vert \tilde \xi \omega \right\vert} < \left
(\frac{r'}{r}\right)^2, \ee where all the dependence on ${\tilde
\xi}$ is negligible in the phase, which reduces to \be (\lambda
{z})\omega \left\vert {\vec r}-{\vec r}^{\prime }\right\vert
\simeq \omega (r-{\hat{n}}\cdot {\vec r}^{\prime }).\ee This
corresponds to the CRAB nebulae case, where the assumptions made
in Ref. \cite{JLM1} are readily verified. A detailed discussion of
synchrotron radiation in the Myers-Pospelov model can be found in
the third reference \cite{UM}.
\subsection{General point of view of LIV radiation}
\label{ss102}
Three paradigmatic examples of Lorentz violating
electrodynamics are given by the effective theories proposed by
Gambini and Pullin
 (GP) \cite{GPED}, Ellis et al. (EMN) \cite{NCST1}, and
Myers and Pospelov (MP) \cite{MP}. They can be written in the
general form of Maxwell equations
\begin{eqnarray}
&&\mathbf{\nabla }\cdot \mathbf{D} =4\pi \rho
,\;\;\;\;\;\;\mathbf{\nabla }\cdot
\mathbf{B}=0,  \label{GENMAXW1}\\
&&\mathbf{\nabla }\times \mathbf{E} =-\frac{\partial \mathbf{B}}{\partial t}%
,\;\;\mathbf{\nabla }\times \mathbf{H}=\;\frac{\partial
\mathbf{D}}{\partial t}\;+4\pi \mathbf{j},  \label{GENMAXW2}
\end{eqnarray}%
with corresponding constitutive relations%
\begin{equation}
\mathbf{D=D(E,B),\;\;\;H=H(E,B)},  \label{GENCR}
\end{equation}%
which we next present in detail for each case, after reviewing the
corresponding equations. Let us recall that the above equations
(\ref{GENMAXW1}) and  (\ref{GENMAXW2}) imply charge conservation $
\partial \rho /\partial t\;+\mathbf{\nabla }\cdot
\mathbf{j}=0, $ independently of the constitutive equations
(\ref{GENCR}). In an abuse of notation we have  denoted by
${\tilde \xi}$ the electromagnetic LIV parameter for all models in
the sequel.
\paragraph{Gambini-Pullin Electrodynamics}
The Maxwell equations for this case are%
\ba &&\mathbf{\nabla \cdot }\mathcal{{B}}=0, \qquad \;\;\;
\mathbf{\nabla }\times \left( \mathcal{E}+2{\tilde \xi}
\mathbf{\nabla }\times \mathcal{E}\right)
+\frac{\partial \mathcal{B}}{\partial t}=0 ,  \\
 &&\mathbf{\nabla \cdot }\mathcal{E}=4 \pi \rho, \qquad \mathbf{\nabla }\times \left( \mathcal{B}+2{\tilde \xi} \mathbf{\nabla
}\times \mathcal{B}\right) -\frac{\partial \mathcal{E}}{\partial
t}=4\pi \mathbf{j},   \label{GP}
\ea%
where the electric and magnetic fields are identified from the
homogeneous equation
as%
\begin{equation}
\mathbf{E}=\mathcal{E}+2{\tilde \xi} \mathbf{\nabla }\times
\mathcal{E},\qquad\mathbf{B}=\mathcal{B}.
\end{equation}%
From the inhomogeneous equations we obtain
\begin{equation}
\mathbf{D}=\mathcal{E}, \qquad \mathbf{H}=\mathcal{B}+2{\tilde
\xi} \mathbf{\nabla }\times \mathcal{B},
\end{equation}%
which  together with the constitutive relations
\begin{equation}
\mathbf{D}+2{\tilde \xi} \mathbf{\nabla }\times \mathbf{D=E}, \qquad \mathbf{H=B}+2{\tilde \xi} \mathbf{%
\nabla \times B},  \label{CRGP}
\end{equation}%
leave the equations in the required form. These equations define
the corresponding functions stated in
 (\ref{GENCR}). In momentum space we have%
\begin{equation}
\mathbf{D}=\frac{1}{1+4{\tilde \xi} ^{2}k^{2}}\left(
\mathbf{E}-2i{\tilde \xi} \mathbf{k }\times \mathbf{E}+4{\tilde
\xi} ^{2} \left( \mathbf{k } \cdot \mathbf{E}\right)\mathbf{k }
\right),\qquad \mathbf{H}=\mathbf{B}+2i{\tilde
\xi}\mathbf{k}\times\mathbf{B}.\label{GPCE}
\end{equation}%
The admixture of vectors and axial vectors in the constitutive
relations precludes the parity violation exhibited by  the model,
together with the presence of birefringence.
\paragraph{Ellis et al. Electrodynamics}
In this case the modified Maxwell equations are \ba &&\nabla \cdot
\mathbf{B}=0, \qquad\qquad\qquad \mathbf{\nabla }\times
\mathbf{E}+\frac{\partial \mathbf{B}}{\partial t}=0,  \\
&&\mathbf{\nabla }\cdot \mathbf{E}+\mathbf{u}\cdot \frac{\partial \mathbf{E}}{%
\partial t}=4\pi \rho _{eff}=4\pi (\rho -\mathbf{u\cdot j}),   \\
&&\mathbf{\nabla }\times \mathbf{B}-\left( 1-\mathbf{u}^{2}\right) \frac{%
\partial \mathbf{E}}{\partial t}+\mathbf{u}\times \frac{\partial \mathbf{B}}{%
\partial t}+ \left( \mathbf{u}\cdot \mathbf{\nabla }\right) \mathbf{E}=4\pi \mathbf{j}_{\, eff}\nonumber \\
&&\qquad\qquad\qquad\quad  =4\pi (\mathbf{j+u(\rho -\mathbf{u\cdot
j})}), \label{ELLIS} \ea which can be written in the form
 (\ref{GENMAXW1}-\ref{GENMAXW2}) via the constitutive relations \cite{NCST1}
\begin{eqnarray}
\mathbf{H} &=&\mathbf{B}-f(\omega)\mathbf{k}\times \mathbf{E},
\quad  \mathbf{D} =\left( 1-f^{2}(\omega)k^2\right)
\mathbf{E}+f^{2}(\omega)\mathbf{k}\left( \mathbf{k}\cdot
\mathbf{E}\right) -f(\omega)\mathbf{k}\times \mathbf{B},\nonumber
\\ \label{CRELLIS}
\end{eqnarray}%
where we have assumed that $\mathbf{u}=f(\omega)\mathbf{k}$ in
momentum space. Taking $\mathbf{u}$ as a vector, this model
conserves parity and shows no birefringence.
\paragraph{Myers-Pospelov Electrodynamics}
This case corresponds to the equations (\ref{MP}). From the last
one we can infer the constitutive relations \be
\mathbf{H}=\mathbf{B}-{\tilde \xi}
\partial _{0}\mathbf{E}, \qquad \mathbf{D}=\mathbf{E-}{\tilde \xi} \partial
_{0}\mathbf{B}, \label{CRMP} \ee which produce
\begin{equation}
\nabla \cdot \mathbf{E=}\nabla \cdot \mathbf{D,}
\end{equation}%
leaving the third Eq. (\ref{MP}) in desired form. Similarly to the
GP case, this model violates parity. In momentum space Eqs.
 (\ref{CRMP}) become \be \mathbf{H}=\mathbf{B}+i{\tilde \xi}\omega
\mathbf{E}, \qquad \mathbf{D}=\mathbf{E}+i{\tilde \xi}\omega
\mathbf{B}. \label{MPCR} \ee \vskip .3cm The above constitutive
relations in the three representative models involve linear
relations among
the fields and can be summarized, in momentum space, as the local relations%
\begin{eqnarray}
D_{i}(\omega ,\mathbf{k})=\alpha _{ij}(\omega ,\mathbf{k})E_{i}(\omega ,%
\mathbf{k})+\rho _{ij}(\omega ,\mathbf{k})B_{j}(\omega ,\mathbf{k}),&&\nonumber  \\
H_{i}(\omega ,\mathbf{k})=\beta _{ij}(\omega ,\mathbf{k})B_{i}(\omega ,%
\mathbf{k})+\sigma _{ij}(\omega ,\mathbf{k})E_{j}(\omega
,\mathbf{k}), \label{CRCOMP}
\end{eqnarray}%
where the corresponding momentum dependent coefficients can be
read from the equations (\ref{GPCE}),\ (\ref{CRELLIS}), and
(\ref{MPCR}). Equations (\ref{CRCOMP}) are the most general
expressions in which any pair of linear constitutive relations can
be ultimately written, which allow to express the fields
$\mathbf{D,H}$ in terms of $\mathbf{E,B}$.
\paragraph{Parameterization of the constitutive relations}
Let us consider Maxwell equations in momentum space%
\begin{eqnarray}
&& \mathbf{k}\cdot \mathbf{B}\left( \omega
,\mathbf{k}\right)=0,\qquad\qquad\quad
\mathbf{k}\times \mathbf{E}\left( \omega ,\mathbf{k}\right) =\omega \mathbf{B%
}\left( \omega ,\mathbf{k}\right), \label{HOM} \\
&&i\mathbf{k}\cdot \mathbf{D}\left( \omega ,\mathbf{k}\right)=4\pi
\rho \left( \omega ,\mathbf{k}\right),\quad i\mathbf{k}\times
\mathbf{H}\left(
\omega ,\mathbf{k}\right) =-i\omega \mathbf{D}\left( \omega ,\mathbf{k}%
\right) \;+4\pi \mathbf{j}\left( \omega
,\mathbf{k}\right).\nonumber \\ \label{INHOM}
\end{eqnarray}
Here we discuss the vacuum situation where the non trivial
constitutive relations arise because \ of LIV effects. Let us take
into account corrections up to second order in the LIV parameter
${\tilde \xi}$ and let us assume that we are in a Lorentz frame
where we demand invariance under rotations. This would correspond
to the rest frame $V^{\mu }=(1,\mathbf{0})$ in the MP model, for
example. We can always go to an arbitrary frame by means of a
passive (observer) Lorentz
transformation. In this way we have the general expressions%
\begin{eqnarray}
\alpha _{ij}=\alpha _{0}\delta _{ij}+i\alpha _{1}{\tilde \xi}
\epsilon _{irj}k_{r}+\alpha _{2}{\tilde \xi} ^{2}k_{i}k_{j}, \quad
\rho _{ij}= \rho _{0}\delta _{ij}+i\rho_{1}{\tilde \xi} \epsilon
_{irj}k_{r}+\rho _{2}{\tilde \xi}
^{2}k_{i}k_{j}, && \nonumber  \\
\beta _{ij} =\beta _{0}\delta _{ij}+i\beta _{1}{\tilde \xi}
\epsilon _{irj}k_{r}+\beta _{2}{\tilde \xi} ^{2}k_{i}k_{j}, \quad
\sigma_{ij}=\sigma_{0}\delta _{ij}+i\sigma _{1}{\tilde \xi}
\epsilon _{irj}k_{r} +\sigma _{2}{\tilde
\xi}^{2}k_{i}k_{j},&&\nonumber \\
\end{eqnarray}%
where $\alpha _{A},\beta _{A},\rho _{A},\sigma _{A}, A=0,1,2$, are
scalar functions depending only upon $\omega$, $k=|\mathbf{k}|$,
and ${\tilde \xi}$. The property $\mathbf{k\cdot B}=0$ sets $\beta
_{2}=\rho _{2}=0$ effectively. In vector notation we then have
\begin{eqnarray}
&& \mathbf{D}=\left(\alpha_0+\alpha _{2}k^2{\tilde \xi}
^{2}\right)\mathbf{E} +\left(\rho_0+i\alpha _{1}\omega {\tilde
\xi}\right)\mathbf{B}+
i\left(\rho _{1}-i\alpha _{2}\omega{\tilde \xi}\right){\tilde \xi} \;\mathbf{k\times B},\label{D} \nonumber \\
&& \mathbf{H}=\left(\sigma_0+\sigma _{2}k^2{\tilde \xi}
^{2}\right)\mathbf{E} +\left(\beta_0+i\sigma _{1}\omega {\tilde
\xi}\right)\mathbf{B} +i\left(\beta _{1}-i\sigma _{2}\omega{\tilde
\xi}\right){\tilde \xi} \;\mathbf{k\times B},\label{H}\nonumber \\
\end{eqnarray}
where we have used the second Eq. (\ref{HOM}) together with $
\left( \mathbf{k\cdot E}\right) \mathbf{k}=\omega \left(
\mathbf{k}\times \mathbf{B}\right) + k^{2}\mathbf{E}
$.
\paragraph{The generalized Maxwell equations \cite{UM1}}
Next we substitute (\ref{H}) in Eqs. (\ref{INHOM})  to obtain the corresponding equations for $\mathbf{E\;}$and $%
\mathbf{B}$. The result is
\begin{eqnarray}
&&i\left( \alpha _{0}+\alpha _{2}k^{2}{\tilde \xi} ^{2}\right) \left( \mathbf{k}\cdot \mathbf{E}%
\right) =4\pi \rho,  \label{INHOM1}\\
&&i\left( \alpha_{0}+\alpha_{2}k^{2}{\tilde \xi}^{2}\right)\omega
\mathbf{E}+i\left[ \beta_{0}+i\left( \sigma _{1}+\rho _{1}\right)
\omega {\tilde \xi} +\alpha _{2}{\tilde \xi} ^{2}\omega
^{2}\right] \mathbf{k}\times \mathbf{B}
\nonumber \\
&&\qquad\qquad\qquad\qquad+i\left[
\left(\sigma_{0}+\rho_0\right)\omega+i\left(
\alpha _{1}\omega^2-\beta_{1}k^{2}\right){\tilde \xi}\right]%
\mathbf{B}= 4\pi \mathbf{j}\left( \omega
,\mathbf{k}\right).\nonumber \\ \label{INHOM2}
\end{eqnarray}%
It is convenient to rewrite the inhomogeneous equations in the
compact form
\begin{eqnarray}
iP\left( \mathbf{k}\cdot \mathbf{E}\right)& =&4\pi \rho,
\label{CINHOM1}\\
i\omega P\, \mathbf{E}+iQ\,\mathbf{k}\times \mathbf{B}+
R\,\mathbf{B}&=&4\pi \mathbf{j}\left( \omega ,\mathbf{k}\right),
\label{CINHOM2}
\end{eqnarray}%
by defining \ba &&P=\alpha _{0} +\alpha _{2}{\tilde \xi}
^{2}k^{2},  \qquad  Q= \beta_{0}+i\left( \sigma _{1}+\rho
_{1}\right) \omega {\tilde \xi} +\alpha _{2}{\tilde \xi}
^{2}\omega ^{2},\nonumber \\
&& R=\left( \beta _{1} k^{2}-\alpha _{1}\omega^2 \right) {\tilde
\xi}+i\left(\sigma_{0} +\rho_0\right)\omega. \label{PQR}
\ea%
Now we have only three independent functions which depend on
$\omega$ and $k$. The  inhomogeneous equations (\ref{CINHOM1}) and
(\ref{CINHOM2}) can be solved by introducing the standard
electromagnetic potentials in the Coulomb gauge. The  radiation
approximation is described in terms of the vector potential ${\vec
A}^\lambda$ only, which will exhibit polarizations $\lambda=\pm 1$
in the general case and that satisfies the equation
\begin{equation}
\left[ Qk^{2}-P\omega ^{2}+\lambda  k\,R%
\right] \mathbf{A}^\lambda =4\pi \mathbf{j}_{T}^{\lambda}.
\label{EQAL}
\end{equation}
Here $\mathbf{j}_{T}^{\lambda}$ is the transverse part of the
polarized current. The Green function $G^\lambda(\omega, k)$
corresponding to the operator in the LHS of (\ref{EQAL}) satisfies
in general  a cubic equation in $k$, which determines the
dispersion relations for the propagating photon. This means that
we have three poles $k_\lambda(\omega)$. In fact one of the poles
is due to the $\beta_1$ factor in the expression for $R$ in Eq.
(\ref{PQR}) and it is located  at $k\sim \xi^{-1}$. Therefore its
contribution can be neglected in our effective theory valid for
$k\ll \xi^{-1}$. The remaining two poles will deviate little from
their position when $\tilde \xi=0$ and will allow us to
characterize the propagation mode corresponding to each
polarization $\lambda$ by a refraction index $n_\lambda(\omega)$
to be read from the appropriate dispersion relation in such a way
that $n_\lambda(\omega)={k}_\lambda(\omega)/\omega$. The general
form of the polarized Green function will be
\begin{equation}
G^{\lambda}(\omega,{\mathbf r})=\frac{1}{4\pi r}F^{\lambda}
(\omega) e^{i\omega n_{\lambda}\left(  \omega\right)r}\label{nbr}.
\label{POLGF}
\end{equation}
Notice that from the birefringent case we can go to the
non-birefringent one by taking $
n_{+}(\omega)=n_{-}(\omega)=n(\omega)$, in which case  $
F^{+}(\omega)=F^{-}(\omega)=F(\omega)$.

In the following we make explicit some general properties that
must be satisfied by any   electrodynamics characterized by a
Green function of the type (\ref{POLGF}). The reality of the
electric and magnetic fields  leads to the condition
\begin{equation}
[G^{+}\left(  \omega,{\mathbf r}\right)]^{\ast}=G^{-}\left(
-\omega,{\mathbf r}\right).
\end{equation}
{ This implies the relations}
\begin{equation}
n_{+}^{\ast}(\omega)=n_{-}(-\omega), \qquad
[F^{+}(\omega)]^{\ast}=F^{-}(-\omega). \label{REALTC}
\end{equation}
For a birefringent medium the real and imaginary parts of the
refraction index for circular polarization components can contain
both $\omega$-even and $\omega$-odd terms, provided that they
satisfy the first Eq. (\ref{REALTC}). In the case of a
non-birefringent medium the real part of the refraction index must
be even in $\omega$, while the imaginary part must be odd. We can
see that the refraction indices for the Myers-Pospelov theory, Eq.
(\ref{REFIND}), satisfy these requirements. A detailed discussion
of this generalized point of view is under preparation
\cite{MUGEN}.

\section*{Acknowledgements}

The author acknowledges C. L{\"a}mmerzhal for his invitation to
participate in the {\it ${\rm 339^{th}}$ WE-Heraeus-Seminar:
Special Relativity, will it survive the next 100 years?}, as well
as for his wonderful hospitality at Potsdam. Partial support from
the projects CONACYT-M\'exico-40745-F and DGAPA-UNAM-IN104503-3;
as well as the hospitality of the George P. and Cynthia W.
Mitchell Institute for Fundamental Physics, Texas A\&M  University
are also gratefully acknowledged.

\printindex
\end{document}